# Scientists' bounded mobility on the epistemic landscape

Shuang Zhang[1,2#], Feifan Liu[1,2#], Haoxiang Xia[1,2*]

[1] Institute of Systems Engineering, Dalian University of Technology, Dalian 116024 China

[2] Center for Big Data and Intelligent Decision-Making, Dalian University of Technology, Dalian 116024 China

* Correspondence to: hxxia@dlut.edu.cn

[#] These authors contribute equally.

**Despite persistent efforts in revealing the temporal patterns in scientific careers, little attention has been paid to the spatial patterns of scientific activities in the knowledge space. Here, drawing on millions of papers in six disciplines, we consider scientists' publication sequence as "walks" on the quantifiable epistemic landscape constructed from large-scale bibliometric corpora by combining embedding and manifold learning algorithms, aiming to reveal the individual research topic dynamics and association between research radius with academic performance, along their careers. Intuitively, the visualization shows the localized and bounded nature of mobile trajectories. We further find that the distributions of scientists' transition radius and transition pace are both left-skewed compared with the results of controlled experiments. Then, we observe the mixed exploration and exploitation pattern and the corresponding strategic trade-off in the research transition, where scientists both deepen their previous research with frequency bias and explore new research with knowledge proximity bias. We further develop a bounded exploration-exploitation (BEE) model to reproduce the observed patterns. Moreover, the association between scientists' research radius and academic performance shows that extensive exploration will not lead to a sustained increase in academic output but a decrease in impact. In addition, we also note that disruptive findings are more derived from an extensive transition, whereas there is a saturation in this association. Our study contributes to the comprehension of the mobility patterns of scientists in the knowledge space, thereby providing significant implications for the development of scientific policy-making.**

## Introduction

Science is a complex system in which a massive population of scientists collectively produce knowledge. The science of science (SciSci) area focuses on studying this system itself[1,2]. This emerging area has been triggering novel research paradigms and providing unique perspectives, covering knowledge dynamics about the emergence of new ideas[3,4], disciplinary evolution[5-7], diffusion of impact[8,9], along with the career dynamics of scientists, including collaboration[10-12], productivity[13], and imapct[14,15], in an attempt to explore the efficiency and laws of scientific development. One important but underappreciated aspect of the system of scientific knowledge production is the scientists' mobility in the knowledge space, which is an abstract space of all scientific knowledge. More specifically, consider the production of knowledge at the scale of an academic discipline; the acquired knowledge, which can be embodied as all scientific publications in the discipline, constitutes a virtual "epistemic landscape" as a subset of the knowledge space. The transition of a scientist's research topics is then related to their "mobility" on this landscape, which is of critical importance for their academic career. The mobility of the massive population of scientists, in turn, has a great impact on the advances of the overall discipline and the volatility of the landscape. For example, the burst of hot topics and integration of knowledge across different sub-fields often associate with the scientists' mobility[16]. Therefore, it is deserved to systematically examine such mobility on the "epistemic landscape" and its interplay with the advances of science.

There have been many studies on scientists' mobility on the epistemic landscape, e.g., scientists' strategies for research problems or topic selection[17-19], research portfolios[20-22], and research topic switching[23-25]. However, one major obstacle of quantitative investigations on scientists' mobility over their scientific careers is the lack of an appropriate quantitative metric for scientists' topic transitions, which hinders the more accurate investigations in this respect. Intuitively, this obstacle can be overcome by developing a model for the epistemic landscape where the distance between two positions is quantitatively measurable, analogous to the geographical maps that can be used to measure the distance between two sites in the physical world. With such a quantifiable epistemic landscape, scientists' mobile trajectories can be tracked, and their mobile patterns can consequently be investigated in a more thorough way. Furthermore, this encourages some key research questions on scientists' mobility patterns, the underlying mechanisms, and the correlation with the academic outputs. First, are scientists' mobility on the epistemic landscape statistically comparable to the albatrosses that wander in a Lévy-flight fashion[26], since research-problem-searching may to some extent be analogous to animal-food-foraging and scientists work collectively to divide labor in an attempt to effectively explore the unknown territory in the knowledge space and avoid occupying the same space simultaneously[27,28]? In the real-world context, would the scientists' topic transitions be more bounded within a relatively narrow range due to the boundaries of knowledge domains and/or the often-observed risk-aversion tendency? Second, can the observed mobility patterns be reasonably explained, especially from a behavioral aspect of the scientists' decision-making on problem selection and topic transition? Third, whether and how would the diversity of scientists' mobility correlate to their academic outputs, in terms of productivity, recognition, and disruptiveness of the researches? Further examinations on these questions are helpful to deepen our understanding of the complex system of science, as well as to facilitate scientific policymaking.

In this work, we use machine-learning techniques to develop a framework to construct quantifiable "epistemic landscapes" at the disciplinary scale. With the constructed epistemic landscapes, we then examine scientists' mobility as an attempt to partially answer the preceding questions. Our endeavor highlights investigating scientists' academic behaviors in the knowledge space comprehensively, thereby enriching studies from a spatial perspective. Our major contributions are as follows. First, we examined the spatial scale by observing the pace and radius of scientists' mobile trajectories, finding the bounded nature of scientists' research topic transitions. Second, we analyzed a hybrid "exploration-exploitation" pattern across the scientists' scientific careers, uncovering their risk-averse strategy that generates the research frequency bias and knowledge proximity bias. Third, we examined the non-straightforward correlation between mobility scope and academic performance, revealing that the scientists' extensive research radius is more associated with disruptive findings, whereas the persistent concentration in the narrower research radius is conducive to bringing higher academic recognition. Our findings are validated by investigating scientists across six large-scale disciplines.

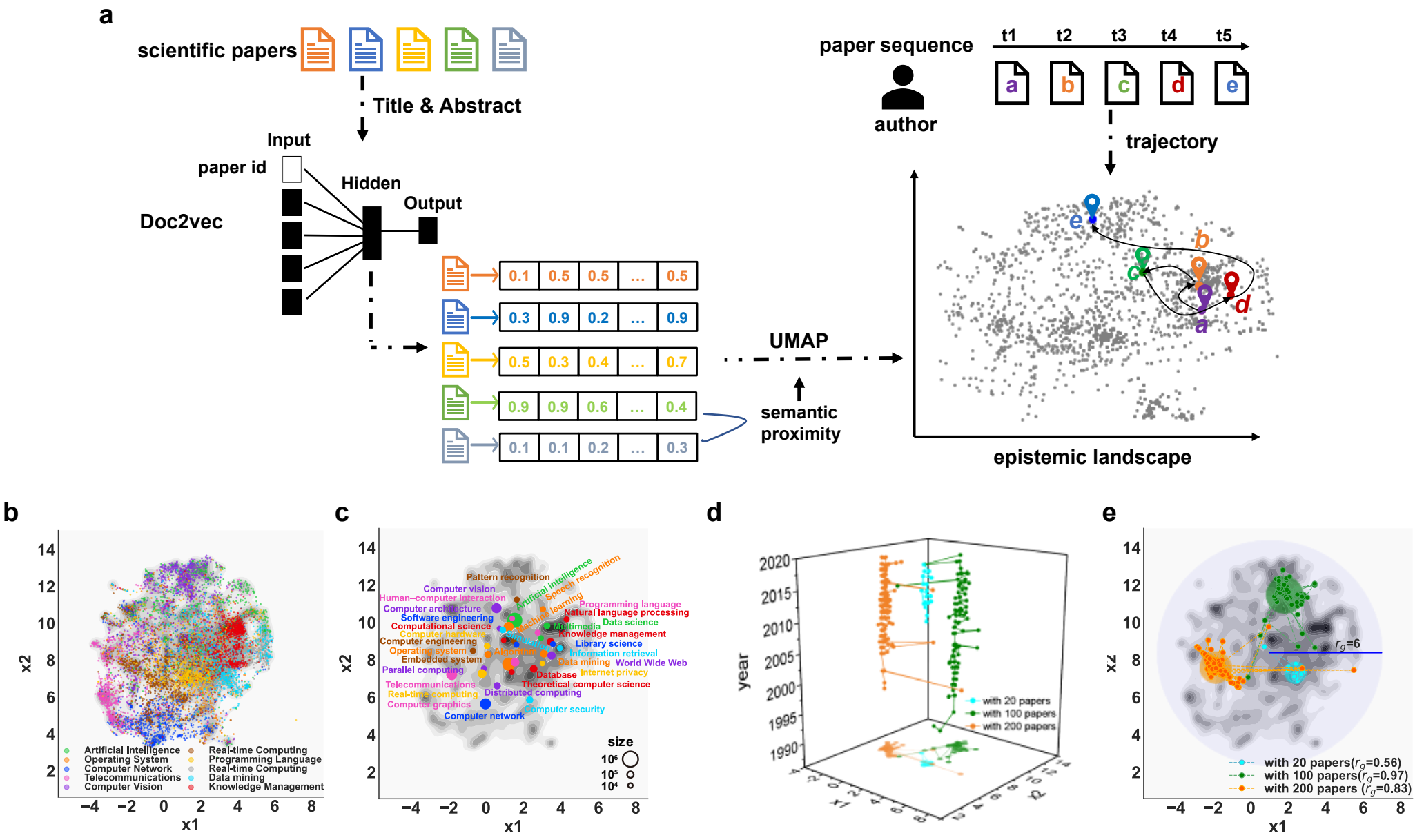

**Fig. 1 | Visualizations of the epistemic landscape and scientists' mobile trajectories.** **a**, Illustration of constructing the epistemic landscape and scientists' mobile trajectories. We first collect scientific papers of a given discipline. Then, we input the preprocessed title and abstract of papers to the document-embedding model Doc2vec. After the training, the high-dimensional semantic vectors of papers are obtained. Then, the manifold learning algorithm UMAP is used to project semantic vectors into the two-dimensional space based on the semantic proximity between vectors calculated by cosine distance. Next, two-dimensional coordinates of each paper are obtained and the global epistemic landscape is hereto generated. Finally, by locating the paper sequence of each scientist, the track of moving trajectories is constructed (see Methods for the detail). **b-c**, The representation-learning-based CS epistemic landscape. The bottom grey shadow corresponds to the Gaussian kernel density of the paper points. In **b**, several paper points sampled are colored with their unique subfield labels. For clarity, the largest ten subfields are presented. The same-colored paper points are clustered together. In **c**, the centroid of each subfield is calculated by averaging the coordinates of the papers in each subfield. The sizes of nodes represent paper counts of each subfield. With the layout of labels, the discipline's knowledge structure can be sketched. **d-e**, Three exemplary mobile trajectories on the CS epistemic landscape. We randomly pick three scientists who have published respectively 20 papers, 100 papers, and 200 papers and track their trajectories using the framework described above. **d** is three trajectories with a year timeline, and **e** is three trajectories on the CS epistemic landscape. In **e**, the radiuses of gyration ($r_g$) of the trajectories represented by colored-shaded circles are small, relative to the radius of the CS landscape ($r_g = 6$). See Methods for calculation of $r_g$

## Results

**Scientists' trajectories on the epistemic landscape.** To investigate scientists' mobility in the knowledge space, we first develop a framework to construct disciplinary epistemic landscapes from bibliometric corpora by combining embedding and manifold learning algorithms, which provides a continuous distance metric and offers a better proxy to track scientists' mobile trajectories.

As illustrated in Fig. 1a, the scientific papers' content vectors are trained by the document-embedding algorithm Doc2Vec[29] and they tensor the domain knowledge space; the high-dimensional space is then projected to a two-dimensional space by the dimensionality-reduction algorithm UMAP[30] based on the semantic proximity of papers. We call the obtained plane the epistemic landscape in line with the concept in the sociology of science[27]. With this "epistemic landscape" being constructed, we obtain the scientists' mobile trajectories by locating their publication sequences. To statistically understand the scientists' virtual mobility behavior, we analyze six research areas, ranging from Computer Science, Physics, Chemistry, Biology, Social Science, to Multidisciplinary Science (see Methods for data description). In the main text we mainly report our results on analyzing the Computer Science (CS) data extracted from Microsoft academic graph (MAG)[31], involving 4,752,206 scientists and 4,391,220 papers from 1958 to 2019. The results of the other five datasets are similar to those of CS and are summarized in Supplementary Notes 4-5.

With the previously-described framework, the constructed epistemic landscape in the CS discipline is exhibited in Fig. 1b,c. First, we attempt to depict the knowledge structure of the epistemic landscape. With the MAG's fields of study classification, Fig. 1b presents an evident clustering structure, and the main subfields in Fig. 1c are semantically distributed. The epistemic landscape is further annotated by key phrases extracted from the text content of papers with the KeyBERT algorithm[32] (Supplementary Fig. S1). The layout of subfields and the concise research topic of the specific region exhibited above depict a clear knowledge structure of this discipline. In Supplementary Note 1, we further systematically validate the constructed epistemic landscape by examining its local and global structure with the citation relationship and classification labels (Supplementary Fig. S2). The repeatedly confirmed validity of the landscape supports the follow-on construction of scientists' mobile trajectories.

Based on the constructed epistemic landscape, we attempt to examine the scientists' mobility patterns. We consider all scientists who have published at least 10 papers as focal scientists, in order to ensure that there are enough track points in their careers for meaningful quantification, resulting in 180,339 focal scientists. By randomly selecting three scientists and taking their trajectories as examples, we observe one's trajectory is locally distributed, albeit with occasional distant track points (Fig. 1d,e). This pattern is different from the Lévy-flight pattern that has been widely observed in the physical world, especially in animal foraging behaviors[33]. Lévy flight, consisting of alternating many local movements and a few long-distance relocations, facilitates rapid dispersal to wider areas at a low cost, and thereby allows greater access to sparsely distributed resources. Yet, long-distance relocations are hardly seen in the visualization of the scientists' trajectories, as exampled in Fig. 1d,e, as well as in various other tested samples. This non-Lévy-flight-like characteristic of scientists' trajectories is worthy of systematic verification to answer the question of how scientists "forage" for research topics within the knowledge space.

**The bounded nature of scientists' mobility.** Following the preceding visualization of the scientists' non-Lévy-flight-like trajectories, we next attempt to quantify the spatiotemporal characteristics of scientists' mobile trajectories on the epistemic landscape. From the temporal aspect, we calculate time intervals ($\Delta t$) between an individual scientist's two successively published papers (Fig. 2a). The fitting results show that $\Delta t$ is better approximated by the truncated power-law than by the power-law. In this study, the tests on heavy-tailed characteristics are performed by the method of Clauset et al[34] and the method of Alstott et al[35], and the statistics are shown in Supplementary Table S1. The truncated exponential tail indicates there is no excessive longer stagnation, which distinguishes scientific publishing activities from various human daily activities (such as sending emails or making phone calls) where waiting times are characterized by heavy tails[36]. Indeed, for scientists' paper-publishing activities, there is an absence of prolonged inactivity.

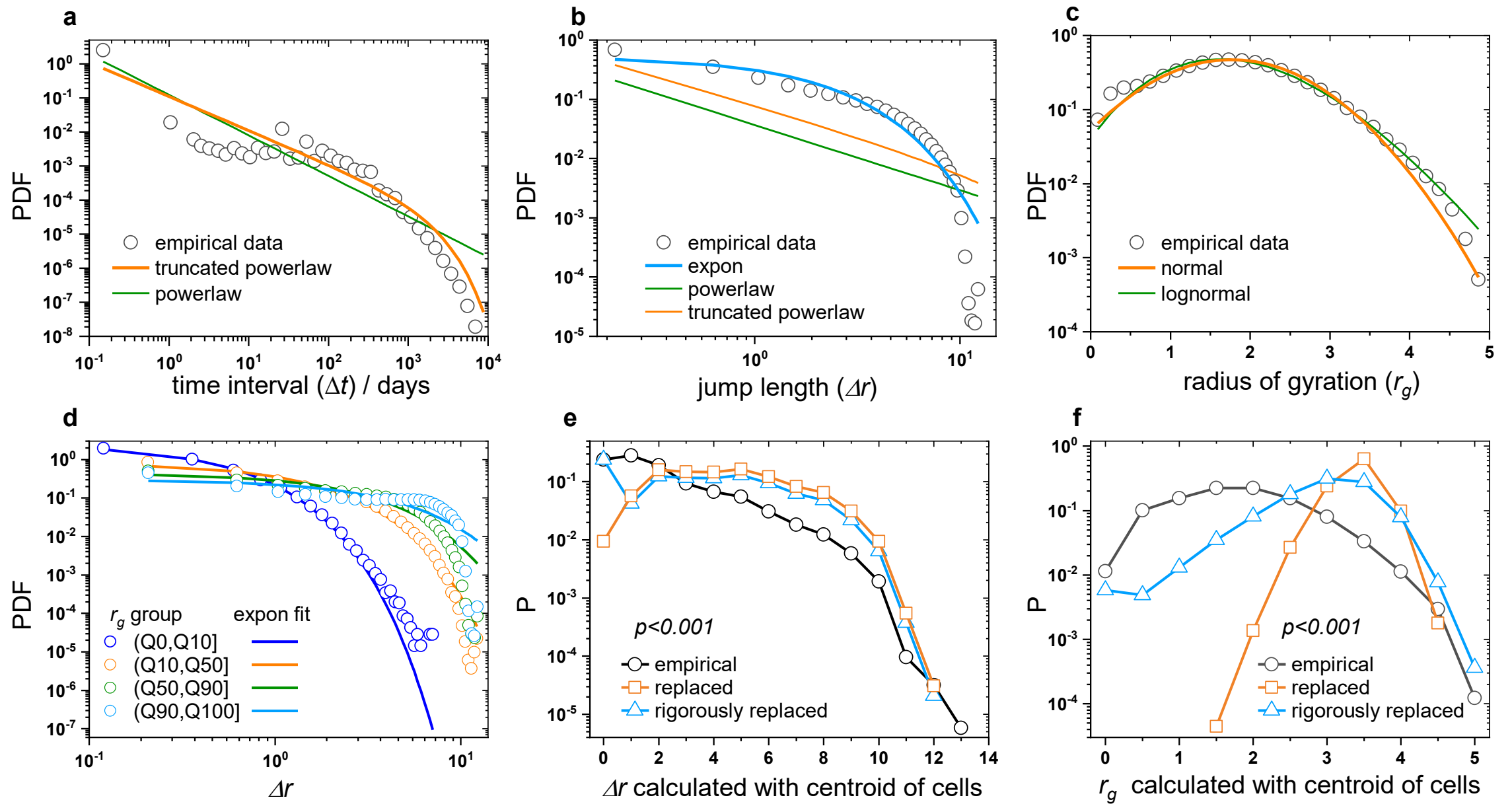

**Fig. 2 | The Non-Lévy-Flight characteristics of scientists' mobile trajectories. a-c**, Empirical results of spatiotemporal characteristics of scientific trajectories, namely time interval (**a**), jump length (**b**), and radius of gyration(**c**). **a**, The time interval of publishing is ($\Delta t$) the time lag between consecutive papers in one's paper sequence, better fitted with the truncated power-law than the power-law. We shift the values by 0.1 (i.e., the minimal value of $\Delta t$ is 0) to make all points visible in the logarithmic scale. **b**, The distribution of the "jump lengths" ($\Delta r$) of each "single move" approximates the exponential rather than the truncated power-law and the power-law. **c**, The distribution of the radius of gyration ($r_g$), referring to the typical distance from an individual to the centroid of mass of their trajectory, is better approximated by the normal or lognormal function. **d**, The distributions of $\Delta r$ for scientists with different $r_g$. By dividing scientists into four groups at their $r_g$ percentile of 0–10, 10–50, 50–90, and 90–100, we find that the four distributions of $\Delta r$ fit well with an exponential function. **a-d**, The comparisons of candidate distributions and statistics are shown in Supplementary Table S1. **e-f**, Comparisons with two controlled experiments. In the replaced case, each original paper in an individual paper sequence is replaced with one randomly drawn from the dataset. The second is a stricter case, where we only replace the distinct cells in the individual trajectory (see Supplementary Note 2 for details). Thus, we generated two kinds of modified trajectories for focal scientists. The $\Delta r$ and $r_g$ are recalculated with the centroid of cells in three cases (i.e., in the real case and two controlled experiments). Both indicators distribute significantly left-skewed in the real case, compared with the results of the controlled experiments. The statistics and P values of the Kolmogorov–Smirnov test are shown in Supplementary Table S2.

From the spatial aspect, we study the spatial scale of the scientists' mobile trajectories with two indicators, namely the jump length and the radius of gyration (detailed calculations are shown in Methods). The jump length ($\Delta r$) is the epistemic distance between one's successive publications. We observe $\Delta r$ is better approximated by the exponential than the power-law function (Fig. 2b). It indicates that scientists' movements are composed of massive short-distanced jumps and a few longer-distanced ones, but the jump distances generally do not scale up. The other indicator, the radius of gyration ($r_g$), is calculated as the typical distance from the center of individual trajectory, depicting the epistemic span of a scientist's publications over a long period (even throughout their academic career). We find that $r_g$ approximates the Gaussian distribution, suggesting the majority concentrate on the narrow and middle knowledge territory and only a few scientists conduct research in a broader landscape (Fig. 2c). From the exponentially distributed $\Delta r$ and Gaussian-distributed overall $r_g$, we observe the non-Lévy-flight characteristics of scientists' mobility. The quantified spatial constraint is also consistent with the visualization in Fig. 1d, e. The observations deviate from human trajectories in physical space, which are mostly characterized by fat-tailed distributions of jump length and radius of gyration[37,38]. Furthermore, we find that the distributions of scientists' $r_g$ in various subfields approximate the Gaussian distribution (Supplementary Fig. S3a). This means that the characteristic of the $r_g$ in the discipline does not arise from the heterogeneity of subfields. Moreover, we observe the movement of scientists is limited in subfields. For subfields, the means of its scientists' $r_g$ are smaller than $r_g$ of its papers distributed. While the larger the knowledge of subfields span, the broader the scientists move (Pearson correlation coefficient is 0.65, Supplementary Fig. S3b). These results quantify the inter-subfields gap in the discipline.

The exponential jump length may result from either each scientist jumping exponentially or a substantial heterogeneity of the whole population, with certain scientists moving within a smaller range and others moving in a larger one. We test these two hypotheses by examining distributions of $\Delta r$ for scientists with a different $r_g$ percentile. In Fig. 2d, it is observed that the exponentially distributed $\Delta r$ is independent of the value of $r_g$. Both scientists with small $r_g$ and those with large $r_g$ move with many short and a few longer jump lengths, simply at variance with the proportion. This reveals that scientists' movements on the epistemic landscape manifest highly spatial localization.

The observed bounded nature of scientists' mobility is further supported, compared with controlled experiments (Fig. 2e,f). We set two kinds of random cases. Beforehand, we mesh the epistemic landscape by $20 \times 20$ and represent individual trajectories with coordinate sequences of cells instead of papers. With this coarse-graining process, it is convenient to determine which area the trajectory points are concentrated in and to design controlled experiments. The first experiment is to replace each scientist's paper sequence with one randomly drawn from the landscape. The second is a more rigorous case, where each distinct cell of one's paper sequence is replaced. In the modified sequence, the visitation frequency, visitation order, and variety of cells are unchanged (see Supplementary Note 2 for the detail). Finally, we recalculate the jump lengths and radiuses of gyration in the controlled experiments and with the center coordinates of cells. The real distributions of $\Delta r$ (Fig. 2e) and $r_g$ (Fig. 2f) are significantly left-skewed against the results of both controlled experiments, indicating that scientists' mobility concentrates on a rigidly restricted territory with relatively small steps.

**Mixed proximity-based exploration and familiarity-based exploitation patterns.** Subsequently, we explore the dynamics of scientists' trajectories on the epistemic landscape. In the previous section, when calculating the $\Delta r$ using the coordinates of cells' track points, we observe that many jumps are sticking in situ both for the real and strict replaced cases (Fig. 2e). More specifically, 24% of the same topics are visited consecutively, which is higher than the 18% in the case where the cell sequence is randomly ordered (Supplementary Fig. S5). This suggests that scientists have certain topic inertia when publishing (i.e., they prefer to publish multiple articles in succession on a focal topic before moving towards another one).

We then examine the visitation frequencies of unique cells by individual scientists, with the scientists being divided into three groups according to the quantile of the number of unique cells ($S$) in their trajectories (Fig. 3a). We find Zipf's law in the topic visitation, independent of $S$. The overall visitation frequency $p$ of the $k$th most visited cell of a scientist is well approximated by $k^{-\zeta}$, where $\zeta \approx 1.2$ is consistent with the coefficient in human mobility[39]. This result suggests that scientists put heterogeneous effort into their research topics, indicating a strong tendency to return to a primary "residence." To further support this observation, we set two kinds of modified trajectories to compare (Fig. 3a). The first is to replace original papers in an individual paper sequence with ones randomly drawn from the dataset. The second is to merely randomize the visitation frequency of each unique cell in one's trajectories (see Supplementary Note 2 for the detail). In the first paper-replaced experiment, Zipf's law does not hold. In the randomization of frequency experiment, the visitation frequency of the most frequently visited cell is lower than that in the real results.

As one's papers are published one by one, individual scientists gradually wander across the epistemic landscape. We find the overall research radius ($r_g$) and the number of visited distinct cells ($S$) present sublinear growth (Fig. 3b,c). After dividing scientists into four groups based on their $r_g$ quantiles, we observe consistent sublinear growths (Supplementary Fig. S6 shows the fittings for these subgroups).

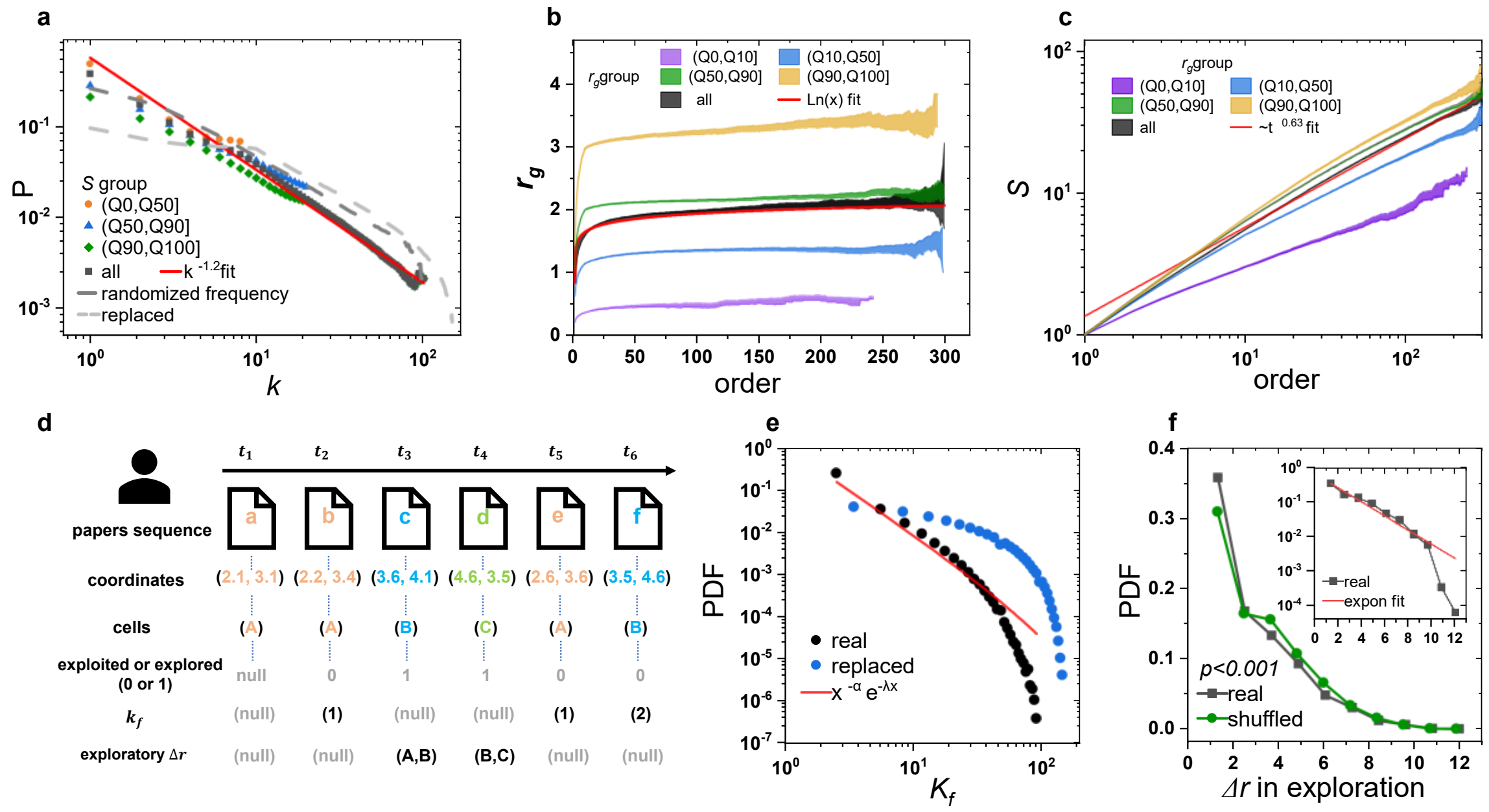

**Fig. 3 | The preference in the mixed exploitation-exploration process. a**, The visitation frequency $p$ of the $k$th most visited cell in each scientist's trajectory. The overall data are well approximated by $k^{-1.2}$ ($R^2 = 0.989$). $S$ represents the number of unique cells that appear in one's trajectories. We divide scientists into three groups at the $S$ quantile of 0–50, 50–90, and 90–100, and observe similar phenomena across groups. Zipf's law disappears in two controlled experiments (dash lines, see Supplementary Note 2 for experimental design). **b-c**, The growth of the $r_g$ and $S$. Bootstrapped 95% confidence intervals are shown as colorful zones. Here, we merely examine scientists with no more than 300 papers (i.e., accounts for 99.8% of the population). We divide scientists into four groups according to the quantile of their $r_g$. With the order of papers in one's publication sequence, the growths of $r_g$ and $S$ are respectively found in line with the logarithmic fitting form $A + Bln(t + C)$ ($R^2 = 0.833$) and exponential fitting function $t^{\alpha}$ ($R^2 = 0.995$), independent with $r_g$. **a-c**, The fitting results for these subgroups are shown in Supplementary Fig. S6. **d**, Illustration of the division of the exploitative and explorative points. Consider a scientist with a trajectory set $= (A_a, A_b, B_c, C_d, A_e, B_f)$, where lowercase letters represent the paper id and uppercase letters represent the cell where the paper point is located; the exploitative points are $(A_b, A_e, B_f)$and the explorative points are $(B_c, C_d)$ in the illustrated case. For example, at $t_5$, this scientist published paper $e$ in cell $A$, which is the cell their previous papers $a$ & $b$ are in; therefore, $A_e$ represents an exploitative track. Furthermore, the frequency rank $K_f$ of each exploited cell is calculated (e.g., the $k_f$ of $(A_b, A_e, B_f)$ is $(1, 1, 2)$). The epistemic distances of the explorative moves (e.g., the point pairs $(A_b, B_c)$ and $(B_c, C_d)$), are calculated. **e**, The $k_f$ of exploitative track points in trajectories present the truncated power-law characteristic, which is absent in the replaced experiment. **f**, The exponentially distributed $\Delta r$ in exploration (**inset** of **f**), with a higher proportion of small $\Delta r$ than that of the shuffled case where the order of explorative tracks is modified (green line, see Supplementary Note 2 for details on modified trajectories). **a**, **e**, **f**, The $P$ values of the KS-test between the distribution of empirical and controlled cases are all less than 0.001, shown in Supplementary Table S2.

The ultraslow diffusive processes reflect the existence of a knowledge boundary, partly supported by scientists' tendency to return to the research areas they have cultivated and explore in small steps with caution. The findings summarized in Fig. 3b,c are consistent with the known individual human mobility patterns in the physical world[39]. Especially, $S$ grows as $t^{\alpha}$, with empirical result $\alpha=0.64$ (Fig. 3c), which is close to the coefficient of human mobility ($\alpha=0.6 \pm 0.02$)[39]. Moreover, the results in Fig. 3a,c coincide with two statistical signatures in the innovation system, namely, Zipf's law[40] and Heap's law[41]. Zipf's law on word frequency and Heaps' law on the growth of distinct words in English documents are well known. In the context of scientists' mobility, the co-existence of Zipf's law and Heap's law suggests that scientists explore new research topics while returning to previous ones. This hybrid process also provides quantitative empirical evidence for Gieryn's study that scientists' problems mostly change gradually rather than abruptly[42]. In other words, there is a simultaneous occurrence of "problem change" and "problem retention," implying a mixed pattern of exploration and exploitation in the research topic transition.

To further detect the preference in the amalgam of explorative and exploitative processes, we divide an individual scientist's track points into the explorative and exploitative parts, depending on whether a specific track point visits a "new" cell in the landscape or revisits a cell that has been visited in the same scientist's previous publications, as illustrated by Fig. 3d. For the exploitative points, we calculate the frequency rank $k_f$ of the corresponding cell, as ref.[43] did; for the explorative points, we measure the epistemic distance ($\Delta r$) between the given point and its preceding point in the scientist's paper sequence. Fig. 3e shows that scientists are likely to return to high-frequency topics in their exploitation activities, with respect to the truncated power-law distributed characteristics of $k_f$. This reveals the existence of the frequency preference in exploitation. Further investigation reveals that about 89% of the returns to the most-visited cell occurred within three steps (Supplementary Fig. S7). For the exploration activities, most of the epistemic jumps are across short knowledge distances, whereas a small number of jumps are with longer distances. The overall epistemic jumps approximate the exponential distribution (Fig. 3f inset), suggesting the existence of proximity preference in scientists' research problem explorations. Moreover, the observed massive short-distanced explorative activities align with Kauffman's notion of "expanding adjacent possible" in innovation systems[44]. But the slight difference in our detection is that scientists also have the intention for long-distance expeditions, which is closely related to scientist's surprisal academic efforts.

To further support these statistical analyses (i.e., truncated power-law distributed $k_f$ and exponential $\Delta r$), we perform comparisons with two controlled experiments. We set replaced trajectories by randomly drawing track points from the dataset and reshuffled trajectories where a given explorative point is replaced by one drawn from the subsequent explorative points in one's trajectory (see Supplementary Note 2 for the details). Then we observe the non-power-law distributed $k_f$ of replaced experiments (Fig. 3e) and a smaller proportion of short explorative jumps in the reshuffled sequence (Fig. 3f), confirming the observation of scientists' frequency bias in exploitation and the proximity preference in exploration.

**The bounded exploration-exploitation model (BEE).** To further probe the mechanism underlying the observed mobility pattern, we propose a bounded exploration-exploitation (BEE) model, as illustrated in Fig. 4a. Different from previous random-walk models[23,24], the proposed BEE model incorporates the scientists' tactical selections in their movements on the epistemic landscape (i.e., the transitions of research topics). On one hand, a knowledge proximity mechanism is included in the model, regarding the non-accidental occurrence of scientists' longer-distance jumps, as scientists may selectively determine how far to move concerning the knowledge proximity. On the other hand, as we previously observed, scientists' preferences in their research histories may also influence the choices of future research topics. This historical memory effect is also considered in the proposed model.

With the above two key mechanisms, the basic setup of agents and the simulation procedure are briefly summarized as follows. The epistemic landscape is represented by a two-dimensional lattice, and scientists take two types of walks on it (i.e., exploring the new and exploiting the previous). Each scientist is initially situated in a cell in the lattice, assigned a specific number of moves, and associated with an individual exploration probability derived from the proportion of the explorative points in the actual trajectories of scientists (Supplementary Fig. S8). Scientist $i$, with exploration probability $p_i$, executes the exploration and picks a new topic under the selection probability that decays exponentially with the increased epistemic distance between the new topic and the current one. Otherwise, with probability $1-p_i$, the scientist executes exploitation and the revisited topic is dependent on the visitation frequency of topics in this scientist's historical trajectory. In addition, we set up two null models for comparison. The knowledge-proximity-based topic selection is neglected in the first null model so that the next topic is randomly selected in both exploration and exploitation. In the second null model, scientists move solely based on exponentially distributed step lengths.

The simulated results by our simple BEE model approximate the real results well, significantly better than those by the two null models (Fig. 4b-f). The significance of differences measured by Wasserstein distance and MAPE are shown in Supplementary Table S3. In the first null model, without the constraint of proximity preference, both $\Delta r$ and $r_g$ are significantly larger than those in the real case (Fig. 4b,c), and the diffusion of $r_g$ is broader (Fig. 4e).

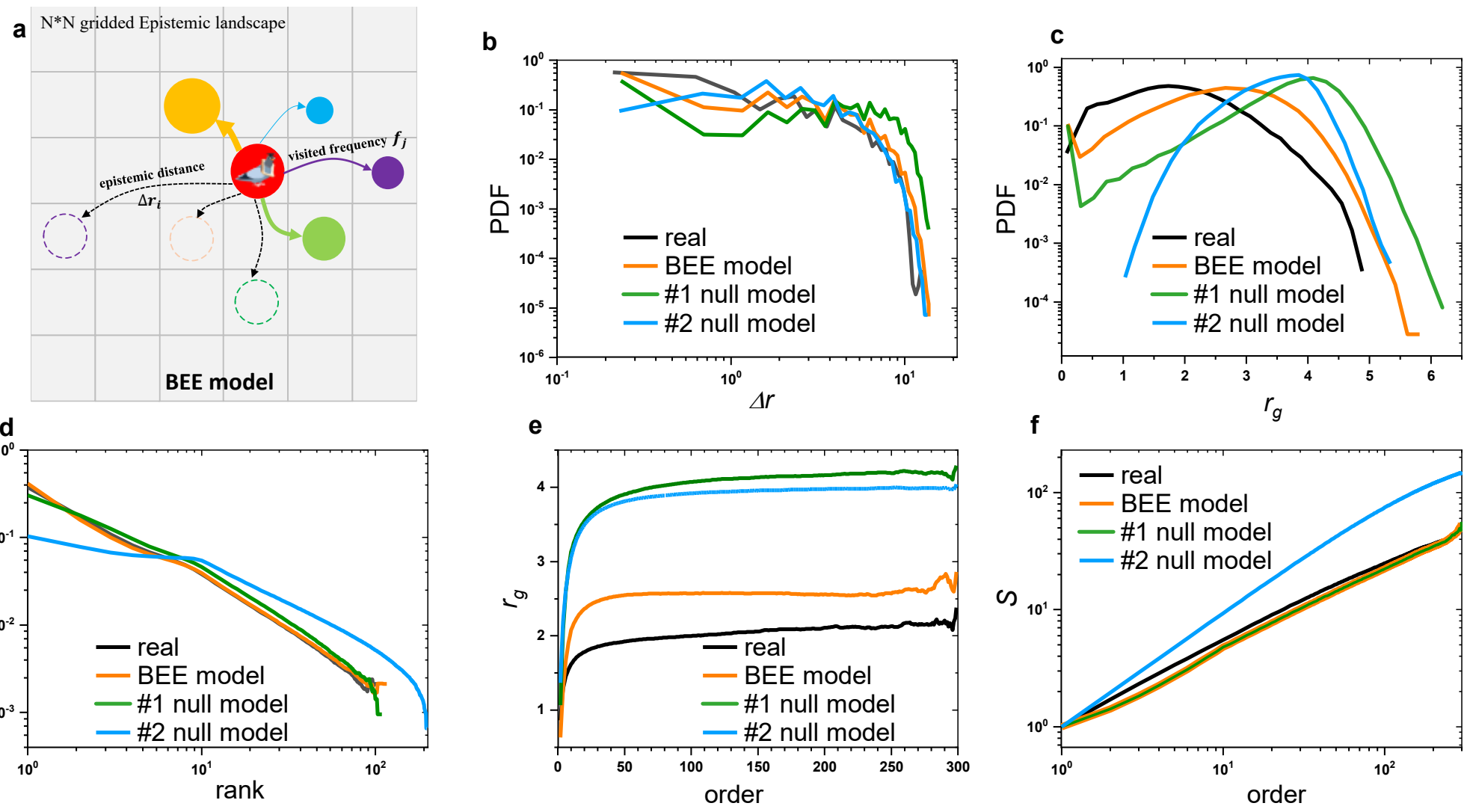

**Fig. 4 | The simulation results of BEE model and two null models. a**, Illustration of the BEE model. Scientists are modeled as selecting cells on one epistemic landscape represented by the gray-gridded plane. Scientists have two sorts of moving selections: exploration and exploitation. There exists frequency and proximity biases in selection. In the exploitation process (solid lines), a scientist determines the next topic based on the frequency of previous visits (the width of solid lines) on each cell. In the exploration process (dotted lines), scientists determine the next topic based on the epistemic distance between a new topic and the current topic (length of the dotted line). We also set two other null models. The first null model (labeled as "#1 null model" in the plots) only contains the return and exploration mechanism. In the second null model (labeled as " #2 null model" in the plots), the agent makes jumps according to the distance, with jump probabilities decreasing exponentially as increasing distance. **b-f**, Comparison between the real and simulated results generated by the BEE model and the two null models. We focus on the distribution of the jump lengths $\Delta r$ (**b**), radius of gyration $r_g$ (**c**), visitation frequency of unique cells $S$ (**d**), increase of $r_g$ as the published order of papers in one's publication sequence(**e**), and increase of $S$ as the published order of papers in one's publication sequence (**f**). In **d-f**, the curves represent mean values. The BEE model better reproduces the observed pattern compared with the two null models. The detailed difference between the simulated results and the actual result is recorded in Supplementary Table S3.

However, there remains a degree of Zipf's law in the visitation frequency, which is derived from the return mechanism (Fig. 4d). The BEE model and the first null model perform consistently in stimulating the growth of $S$, attributing to the same set of the exploration probability (Fig. 4f). In the second null model, the long-distance part of the jump distribution fits the actual data well (Fig. 4b). This further confirms that there are non-accidental long-distance transfers in scientists' mobility. But due to the lack of the return mechanism in the second null model, the small-distance jumps are partly missing (Fig. 4c), Zipf's law is absent (Fig. 4d), and growths of $r_g$ and $S$ are faster than those in the real case (Fig. 4e,f).

Furthermore, as shown in Fig. 5, the individual trajectories generated by the BEE model aggregate to form a hotspot distribution (i.e., the density heat map of track points) that fits the actual data in the discipline (Fig. 5a,b). Here, we additionally set up two lattice-based random-walk models, adapted from previous studies to model knowledge discovery processes[23,24], namely a simple random-walk model and an exploiting random-walk model. In both models, an agent walks on the lattice by jumping arbitrarily to one of the four direct neighbors with equal probability, without the longer-distanced jumping with decaying probability used in our BEE model. The major difference between the simple and exploiting random-walk models is that the latter incorporates the return mechanism and the former does not. Through a comprehensive comparison of these six panels in Fig. 5, in terms of both the range of track point density and the positional distribution of aggregated hotspots, we find that the BEE model produces a more similar visitation diversity of cells to the original. In Fig. 5d,e, the range of density is relatively small, and the higher hotness is clustered in the center of the grid. However, the shape of the distribution of relative hotspots in Fig. 5c,f is closer to the actual case (Fig. 5a) than those in Fig. 5d,e, suggesting that the return mechanism plays an important role in the generation of hotspot positional heterogeneity. However, the range of density in Fig. 5c is relatively small. The BEE model may better capture the scientists' mobility pattern in terms of the aggregate track points in the epistemic landscape of an academic field. The advantage of the BEE model may be explained by the quantification of the continuous knowledge distance for topic transitions or epistemic jumps, and the corresponding combination of frequency-based exploitation and distance-based exploration in scientists' research topic foraging.

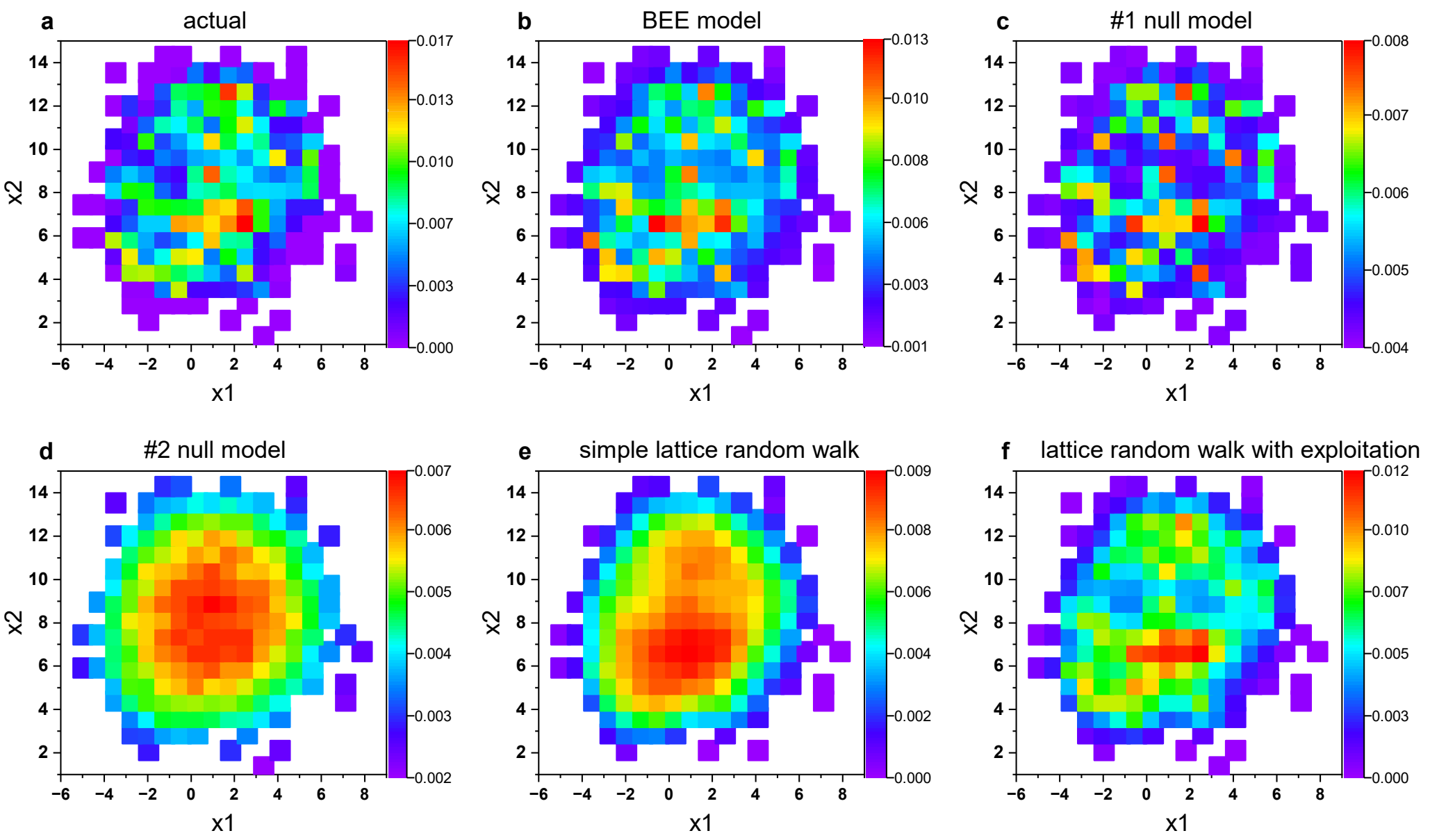

**Fig. 5 | The generated heat maps of scientists' real and simulated track points. a,** generated by the actual scientists' trajectories. **b,** generated by the BEE model. **c,** generated by the #1 null model, which only contains the return and exploration mechanism. **d,** generated by the #2 null model where agents move based on exponentially distributed step lengths. **e,** generated by the simple lattice random-walk model, assuming neighboring lattice jump with equal probability. **f,** generated by the exploiting lattice random-walk model (i.e., adding an exploitation process to the simple random walk). The grid represents the epistemic landscape. By aggregating all trajectories, we calculate the probabilities of the track points distributed on each cell. The redder the cell, the higher the density of track points in that cell. The performance of six models could be evaluated in terms of the scales of heterogeneous probability of distribution of track points and the locational distribution of the hotspots. Of all the simulation results, the BEE model shows the closest pattern to the actual data.

**The non-straightforward correlation between mobility scope and academic performance.** Based on the previous examinations on scientists' bounded mobility, we subsequently investigate how a scientist's radius of gyration ($r_g$), as an indicator of the mobility scope on the epistemic landscape, associates with their academic productivity and recognition as measured by three widely used bibliometric indicators, namely the total count of published papers ($p_c$), average count of citations per paper ($c_c$), and h-index ($h$). By averaging the d-score[45] of all their papers, we also measure the scientist's disruptive value ( $d$ ), which emphasizes the originality and disruption of the individual scientist's research. Fig. 6 shows that the extent of the individual scientist's mobility scope does not have a simple correlation with their academic performance (regarding the four different indicators). For disruptive output, by and large, we can see a steady increase of $d$ (with respect to both percentile in Fig. 6a and relative ratio in Fig. 6b) with the increase of $r_g$, although the increasing slope goes down at the right end of $r_g$. This implies that scientists who are across greater epistemic spans are likely to associate with disruptive studies. This correlational pattern is to some extent similar to that between $r_g$ and the paper count $p_c$, as greater $r_g$ generally associates with greater $p_c$. However, it could be seen that $p_c$ increases with the increase of $r_g$, for $r_g < 2$, but the same trend does not hold for greater $r_g$. For recognition measured by the average count of citations per paper, the correlational pattern is quite different. We can generally observe a decrease of $c_c$ with the increase of $r_g$, except for the reverse increasing trend at the very left end of the $r_g$ coordinate. The previous results show that the narrower mobility range (corresponding to the left end of the $r_g$ coordinate) correlates with lower paper productivity and less disruptive output, but on the other hand, higher academic recognition (represented by a higher count of citations). Oppositely, the broader mobility range (corresponding to the right end of the $r_g$ coordinate) correlates with greater disruptive output and lower academic recognition. Finally, scientists with both high productivity and high recognition are more likely to have intermediate research scopes, as indicated by the inverted U-shaped pattern of the $r_g$ correlation.

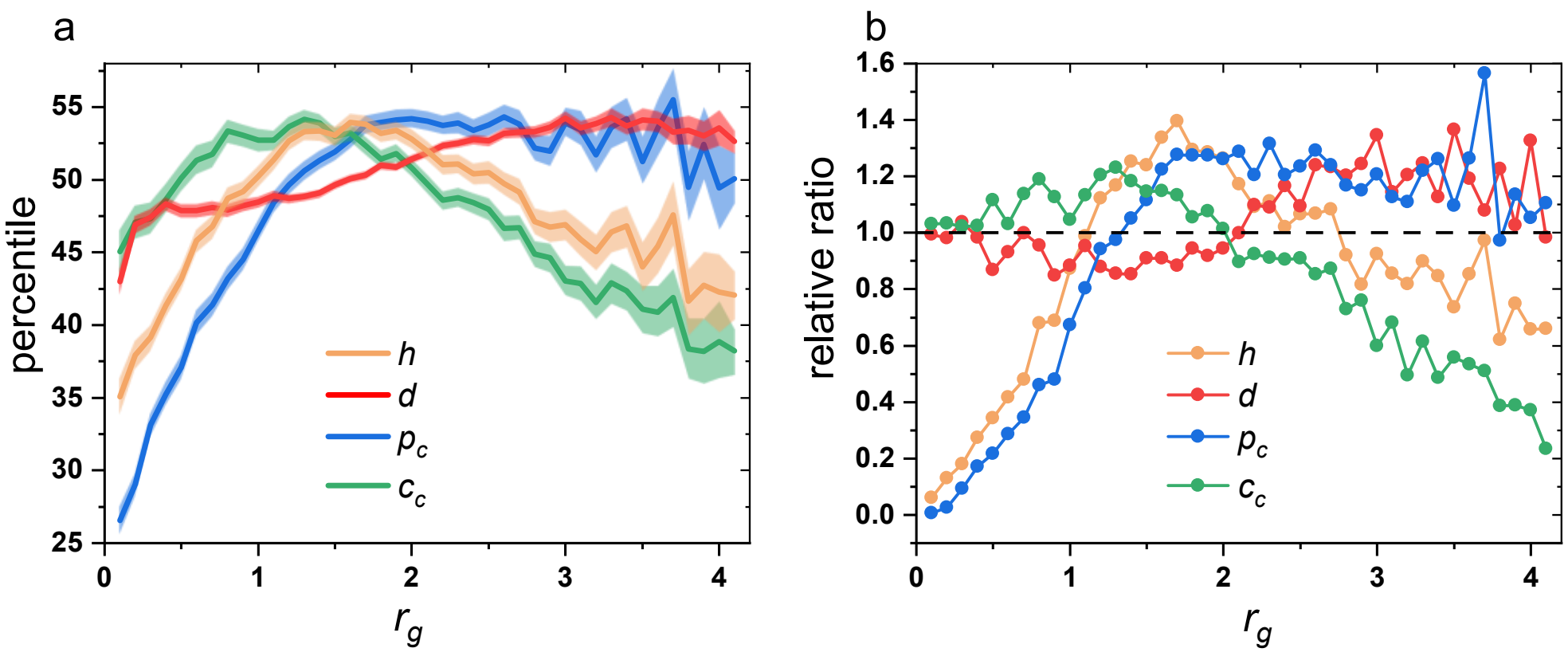

**Fig. 6 | Correlations between the radius of gyration and four key bibliometric indicators.** The four key indicators of an individual scientist's academic performance include the total count of published papers ($p_c$, blue curve), average count of citations per paper ($c_c$, green curve), h-index ($h$, orange curve), and mean disruptive value ($d$, red curve) by averaging the d-score of the individual scientist's papers[45]. **a**, The trend of the mean percentile of bibliometric indicators. Bootstrapped 95% confidence intervals are shown as colorful zones. **b**, The relative ratios of the top 10% of scientists with the highest performance distributed in different $r_g$ groups. After finding the top 10% of scientists with the highest performance under each indicator, we calculate the relative ratio by comparing the actual proportion of these scientists in each group to the ideal value (0.1). In this way, it could be known into which $r_g$ group the top 10% of scientists are more distributed by comparing the relative ratio with the baseline (grey dash line $ratio = 1$). **a-b**, For both average percentile and extreme cases, it is observed that with the increase of $r_g$, $h$ and $c_c$ first increase and then decrease. As $r_g$ increases, $p_c$ incrementally increases, then remains almost constant. As $r_g$ increases, it is a slowly increasing trend of $d$ with a saturation.

**Robustness of results.** We test the robustness of our findings in terms of fine-tuning the algorithm parameters, deploying different embedding algorithms, and extending analysis across different disciplines. The epistemic landscape is the foundation for the quantification of scientists' moves in the knowledge space and the performance of machine learning algorithms often relies on the fine-tuning of parameters. Thus, we examine the interference of the visualization dimension and other key parameter settings of algorithms on the results in Supplementary Note 3 and find the insensitivity of key observations.

In Supplementary Note 4-5, we validate our findings by investigating scientists in other five large-scale disciplines, including Physics, Chemistry, Biology, Social Science, and Multidisciplinary Science. The key findings of spatial characteristics in these six disciplines are consistent. We both observe non-fat-tail characteristics for the jump length and radius of gyration of mobility, and robust simulation results of the BEE model across six disciplines. In terms of the association between $r_g$ and academic performance, consistent conclusions can be drawn from similar patterns presented in all six examined disciplines, despite the slight differences in the trend curves. That is, narrower movement corresponds to higher academic recognition, and scientists whose mobility is in an intermediate area are more likely to achieve higher output and h-index, whereas broader mobility favors the discovery of breakthrough work. Notably, to verify whether the experiment results are dependent on the construction method of the epistemic landscape, we examine the Physics disciplines with a graph embedding algorithm. With the PACS codes provided by APS data, the Physics epistemic landscape is constructed by Node2Vec[46] and UMAP based on the PACS code relational proximity in a co-occurrence network. Although the technique is quite different, the epistemic landscape of Physics in Supplementary Fig. S19 clearly presents the structure of research subfields ranging from electromagnetism, and condensed matters to nuclear physics. And further, in Supplementary Fig. S20-S23, by visualizing physicists' trajectories, quantifying spatial characteristics of trajectories, and examining the mobile preference in the mixed exploitation-exploration process, the applicability of the BEE model and the correlation between activity radius and academic performance, we also observe a similar pattern with those in other disciplines. In a word, we reach the same conclusions across different disciplines and representation learning methods.

## Discussion

An empirically detailed investigation of the regularity of scientists' mobility on the epistemic landscape is an indispensable matter for enlightening scientific innovation activities or guiding policymakers. In our study, based on the embedding algorithms and manifold learning methods, we construct the epistemic landscape with continuous knowledge distance at the disciplinary scale and track scientists' mobile trajectories embodied in publication sequences. In this way, we quantify the spatiotemporal characteristics of scientists' mobile trajectories, explore the mechanism underlying the mobility pattern, and reveal its correlation with their academic performance. Our contributions can further be discussed from three perspectives, including scientists' research topic transitions, human mobility patterns in the knowledge space, and insights into scientific policymaking.

First, this study contributes to deepening the understanding of scientists' research topic transitions in the spatiotemporal level. Apart from quantifying the extent and frequency of research topic change and diversification strategies, here, we put a particular emphasis on the research topic dynamics over a scientist's academic career through regarding the sequential process of topic transition as a mobility process on the epistemic landscape. We empirically observe the scientist's overall research scopes commonly cover a middle range of "territory" of the disciplinary epistemic landscape, whereas most of their topic transitions were combined with many small epistemic and a few long-distance jumps. The bounded nature of topic transition over a scientist's academic career supports the existence of "essential tension" in scientific inquiries[47]. We further find that scientists adopt a blended exploration-exploitation strategy, differing from the previous studies that commonly treated exploration and exploitation as opposites and attempted to distinguish between explorers and diggers[23,25], or explored the optimal choice or combination of strategies at different career stages[21]. Specifically, scientists use frequency and proximity preferences to buffer the tension. We argue that the existence of the frequency bias in problem retention maximally makes scientists guarantee the output with minimal effort, relying on knowledge familiarity, and the proximity bias in problem change demonstrates an extreme risk-averse tendency when seeking originality. This subtle balancing pattern evokes the idea of contextual ambidexterity mentioned in the context of organizational management[48]. Additionally, the BEE model we proposed further confirms the observed pattern and better captures the key characteristics of individual topic transitions than the common topic transition models.

Second, our study extends the understanding of human mobility in the physical and cyber space to the knowledge space. This study recognizes similar regularities between human mobility in the physical world and scientists' mobility on the epistemic landscape, as both types of mobility can be modeled as a return-exploration process with frequency preference, and spatially both present an ultra-slow diffusion process. The key difference lies in the spatial scale. The direct observations in human mobility show the heavy tail for both the jump length and the radius of gyration. However, we observe the characteristic spatial scales of scientists' mobility. A recent study has shown that the scale-free features of human mobility arise from mixtures of normal (or lognormal) distributions across variant geographical scales, such as a block, city, and country[49]. For knowledge space, it is worth investigating whether the overall distributions across the whole science landscape will show the scale-free characteristics. Many studies have also looked at human mobility in cyberspace, including in online games[50], forums[51], and the Websphere[52]. For virtual spaceless environments, based on visual networks describing the transit from one website to another, the scaling law in the distribution of dwelling time and the heterogeneous rank distributions are checked. But due to the lack of metric space, the distributions of displacement and radius of gyration are unable to realize effective measures. Our study complements the mentioned missing factors and provides a reference method for subsequent research on online communities.

Third, our findings also provide insights into scientific policymaking. This study points out the risk-averse tradeoff in scientists' problem selection and transition. Scientists mostly conduct research within an intermediate-scale area and search for new nearby to achieve a higher output and impact. However, our association analysis finds that the limited research transition does not yield higher output, whereas the explorative research transition attenuates the academic impact but increases the probability to produce more disruptive outcomes. The pursuit of scientific breakthroughs requires scientists to take wider knowledge spanning accompanied by greater risks and costs, which calls for guidance and support from scientific policies and funds. These suggest that the scientific evaluation systems should loosen the ties and shift the focus away from the number and citation of publications to make scientists sail at ease. Moreover, excessive knowledge spanning creates excessive breakthroughs, implying the intensity of promoting knowledge crossover should be limited rather than unlimited because excessive diversity often requires excessive fund costs but does not bring as much disruptive efforts.

Several promising research extensions can be performed based on the current work. We only consider the behavioral characteristics of historical trajectories. It is worthwhile to further examine how the observed and modeled behavioral characteristics correlate with various previous analyses on the tradeoffs and contradictions in research topic selection and transitions, such as long-term or short-term returns[53], tradition or innovation[17,18], exploration or exploitation[47], and diversification or specialization[54,55]. Furthermore, a straightforward subsequent topic is to examine how external factors, such as the hotness of research topics and social interaction, would influence scientists' mobility on the epistemic landscape. Another interesting research direction would be the collective effect and competition-cooperation gaming interaction in science. The analysis framework is general and applicable to understanding epistemic mobility in various other communities, such as patent development, open-source software development, and social media areas.

## Methods

**Data.** In this paper, we focus on six disciplines, including Computer Science, Physics, Chemistry, Biology, Social Science, and Multidisciplinary Science. Apart from Physics, the data of the other five disciplines are extracted from Microsoft Academic Graph (MAG). Drawing on MAG's "fields of study" classification[31], we extract 4,752,206 authors and 4,391,220 papers associated with the label of "Computer Science", from 1948 to 2019. Eventually, we select and analyze 180,339 scientists in this dataset with at least 10 papers. The Chemistry data contains journal papers labeled "Chemistry" in the MAG dataset, covering 9,568,741 authors and 6,916,260 papers until 2019. We focus on 458,987 scientists in this data with at least 10 papers. The Biology data contains journal papers marked "Biology" in the MAG dataset, involving 9,731,092 authors and 7,157,231 papers. There are 607,443 focal scientists in this data with at least 10 papers. The Social Science data consists of the papers published in SAGE publishing group journals, with 740,196 authors and 765,709 papers, ranging from 1965 to 2019. We analyze 15,304 focal scientists in this data with at least 10 papers. The multidisciplinary scientific data were collected from five representative multidisciplinary journals, including Nature, Science, Proceedings of the National Academy of Sciences, Nature Communications, and Science Advances. The dataset contains 948,180 authors and 562,998 papers, from 1869 to 2019. There are 20,172 focal scientists in this data with at least 10 papers. The Physics dataset used in this study is the APS journal and the author disambiguated data provided by Sinatra et al[14], containing 482,566 papers from 1893–2010 and 236,884 authors. Finally, we analyze 13,720 physicists with at least 16 papers in the APS dataset.

**The epistemic landscape and scientists' trajectories.** As Fig. 1a illustrates, we mainly utilize the semantic proximity embedded in the textual content of papers to construct the epistemic landscape and individual scientists' mobile trajectories on the epistemic landscape with Doc2Vec[29] and UMAP[30] algorithms. Doc2vec, a popular algorithm in document representation learning, is deployed to learn the word frequency, word order, and word semantics of textual content. After building the corpus with the title and abstract of each paper in the dataset collected, we train the Doc2Vec model until convergence and obtain the embedding vector of the specific research content of each paper. UMAP is an advanced dimensionality-reduction technique, outstanding in its ability to maintain the global and local topology of high-dimensional vectors. We use cosine metrics to measure the proximity between document vectors in the original high-dimensional space and use UMAP to project embedding vectors into a two-dimensional plane. Thus, we get the coordinates of all papers, and the epistemic landscape is generated. Finally, by locating the coordinates of the individual scientist's publication sequence on the epistemic landscape, the scientist's movement trajectory can be traced and quantitatively analyzed by coordinate-based computation.

In addition, the epistemic landscape of Physics is constructed based on the relational proximity learned by the graph-embedding method. We first perform node-embedding learning Node2Vec[46] on the PACS code co-occurrence network and then obtain the low-dimensional coordinate of each PACS code with UMAP. Next, the location of each paper is determined by calculating the barycenter of coordinates of its PACS codes, and finally, the trajectories of physicists are tracked based on their publication records.

**The radius of gyration and the jump length.** Two indicators, the radius of gyration and the jump length, are mainly applied to characterize the spatial scale of scientists' mobility. The radius of gyration ($r_g$), calculated as in formula (1), refers to the typical distance from an individual to the centroid of mass of their trajectory. When applied to scientists' movements, $r_g$ represents the core scope of a scientist's scientific research over their scientific career. The jump length ($\Delta r$), as defined in formula (2), measures the distance on the epistemic landscape between two successive papers in the publication sequence of individual scientists, describing the pace of scientists' mobility on the epistemic landscape.

$$r_g = \sqrt{\frac{1}{N}\sum_{i=1}^{N}(\boldsymbol{r}_i - \boldsymbol{r}_{cm})^2}, \text{ where } \boldsymbol{r}_{cm} = \sum_{i=1}^{N}\boldsymbol{r}_i/N \quad (1)$$

$$\Delta r = \boldsymbol{r}_i - \boldsymbol{r}_{i-1} \quad (2)$$

In formula (1) and formula (2), $\boldsymbol{r}_i$ , $\boldsymbol{r}_{i-1}$ are the coordinates of the scientist's $i_{th}$, $(i-1)_{th}$ papers and $\boldsymbol{r}_{cm}$ is the centroid of the coordinates of the individual scientist's $N$ papers on the epistemic landscape.

**Generation of the track-points heatmap.** We generate the hotspot density map of track points on the epistemic landscape based on our proposed BEE model and other controlled simulation settings. According to the cell in which each track point is located, we aggregate the track points of all scientists and calculate the distribution probability of the track points in each cell. Then, the heat map is drawn where color represents the track-point density of the cell on the epistemic landscape.

### Data availability

The MAG data used in this paper was downloaded via the MAG website https://www.microsoft.com/en-us/research/project/microsoft-academic-graph. However, the Microsoft Academic website and underlying APIs have been retired in 2021. The APS data are available at https://journals.aps.org/datasets by submitting a request.

### Code availability

Code that supports the main findings of this study has been deposited in the GitHub repository https://github.com/cs-dlut/ScholarMobility. All other codes used in this study are available from the corresponding author upon reasonable request.

### Acknowledgments

This work is supported by the National Natural Science Foundation of China under Grant Nos. 72371052 and 71871042 (to HX), and by the Humanities and Social Science Project of the Ministry of Education of China Grant No 18YJA630118 (to HX).

### Author contributions

H.X. and F.L. conceived the study. F.L. and S.Z. designed the research. S.Z. and F.L. performed the experiments. All authors contributed to the analysis of the results and writing of the manuscript.

### Competing interests

The authors declare no competing interests.

Supplementary information for

# Scientists' bounded mobility on the epistemic landscape

Shuang Zhang[1,2#], Feifan Liu[1,2#], Haoxiang Xia[1,2]*

[1] Institute of Systems Engineering, Dalian University of Technology, Dalian 116024 China

[2] Center for Big Data and Intelligent Decision-Making, Dalian University of Technology, Dalian 116024 China

* Correspondence to: hxxia@dlut.edu.cn

# These authors contribute equally.

**This document includes:**

Supplementary Note 1: Validation of the epistemic landscape

Supplementary Note 2: Description of controlled experiments

Supplementary Note 3: Key results under different parameter settings of algorithms

Supplementary Note 4: Extended experiments with other four large-scale disciplines

Supplementary Note 5: Extended experiments in Physics with the graph embedding method

Supplementary Figures S1-S23

Supplementary Tables S1-S4

## Supplementary Note 1: Validation of the epistemic landscape

Our examinations on the scientists' mobility in the knowledge space are based on the construction of a quantitatively measurable epistemic landscape by using machine learning techniques. It is indispensable to test the validity of the constructed epistemic landscape. We mainly utilize citation relations and the fields of study (FoS) classification provided by Microsoft Academic Graph (MAG)[1]. MAG, leveraging both text and graph structure, builds a hierarchy of academic fields, and each publication is tagged with multiple FoS. Here, we mainly use Level-1 (L1) FoS in the hierarchical classification system. There are 34 L1-FoS labels under the L0 "Computer Science" (CS) label. For each paper, we determine its unique L1 FoS as its subfield. Firstly, according to parent–child relationships provided by MAG, we transform the multiple original FoS labels of each paper into multiple L1 FoS labels. Then, for each paper, we sum the confidence score of each L1 FoS and take those corresponds to the maximum value as its subfield.

First, we validate the constructed epistemic landscape by examining whether it exhibits strong clustering of semantically similar papers. Fig. 1b of the main text shows the aggregation of the papers with the same L1 labels. In Fig. 1c, the location distribution of the coordinates of the labels obtained by aggregating the coordinates of their papers clearly divides the map into three knowledge parts. Then we annotate the regions on the epistemic landscape based on the keyword-extraction method. The method of KeyBERT[2] is used for keyword extraction, which extracts the keywords of the paper from its title and abstract. In Supplementary Fig. S1, several marked areas are mainly where the paper density is higher. The key phrase clouds present specific topic contents, in line with FoS to a large extent. These results show that the CS epistemic landscape depicts a clear knowledge structure.

We then mesh the epistemic landscape $20 \times 20$ for systematic analysis. Moreover, as a basis for comparison, we set up a controlled experiment by randomly distorting the position of papers. Firstly, we draw the word cloud of each cell of the epistemic landscape using the multiple FoS labels of all papers located in each cell. We notice the apparent semantics relatedness of the high-frequency keywords in the same word cloud. Due to space limitations, we put the word clouds in the following link: https://github.com/cs-dlut/ScholarMobility. Further, to verify papers in the same cell indeed share the same FoS labels, we measure the purity of FoS distributed in each cell with the entropy index[3],

$$normalized\ entropy = -\sum_j^n \frac{p_j \log_2 p_j}{\log_2 n},$$ where $p_j$ is the frequency of $j$ FoS in one cell.

A smaller normalized entropy indicates a more heterogeneous FoS distribution. Supplementary Fig. S2a shows that both entropies are smaller than that of the controlled experiment, suggesting that a higher fraction of papers in each cell share the same FoS labels. In addition, we construct a direct citation network of the papers and divide the citation communities with the Louvain algorithm[4]. The purity of the distribution of citation communities in one cell is also measured by the entropy index, and the same results are observed, as shown in Supplementary Fig. S2a. These results provide extra evidence that papers on similar topics are clustered on the epistemic landscape and vice versa. Next, we quantitatively check whether the epistemic distance on the landscape can reflect the extent of research dissimilarity between papers. We construct the FoS vector for each cell with the FoS labels located in it and calculate the cosine distance for each pair of cells. As shown in Supplementary Fig. S2b, there is a significantly positive correlation between the content distance and epistemic distance. The correlation coefficient is 0.67, which decreases significantly to 0.21 in random cases.

In Supplementary Fig. S11-S14, we use the proposed methodological framework to obtain epistemic landscapes for large-scale Chemistry, Biology, Social Science and Multidisciplinary Science disciplines. Their knowledge structure is also clearly identified by KeyBERT words. For example, one the epistemic landscape of the Multidisciplinary Science discipline, the several research topic clusters are scattered on the edge, which may reflect the fact that the multidisciplinary data were obtained from journals with very different research contents. We provide additional tests of the applicability of the method in disciplines ranging in size from thousands to millions of papers. Three additional epistemic landscapes generated by other disciplines, and the associated systematic tests of local and global structures, are available at the following link: https://github.com/cs-dlut/ScholarMobility.

The above shows the validity of the constructed epistemic landscape, allowing us to reconstruct the scientists' mobile trajectories. The epistemic landscape captures the semantics of papers' contents, which have a certain correspondence with the citation network's community structure. The papers in the same cell have similar topics, and the epistemic distances could represent the degree of content topical divergence among papers. In short, the epistemic landscape effectively characterizes topical associations at a finer-grained semantic level compared to keywords and it provides the basis for tracing scientists' mobile trajectories in the knowledge space.

## Supplementary Note 2: Description of controlled experiments

In addition to observing the statistical distribution, several controlled experiments are designed to discover the underlying patterns in scientists' mobility. To better implement the experiment before that, we start by meshing the map. All papers are located in the $20 \times 20$ grid. Thus, besides its own coordinates, each paper has the coordinates of the centroid of its cell. As shown in the left panel of Supplementary Fig. S4, the node color indicates the cell to which the paper belongs. To illustrate, consider a particular scientist with a publication record set $P = (a, b, c, d, e, f, g, h, I)$. As shown in the right panel of Supplementary Fig. S4, the letter indicates one paper and the color of a circle indicates the cell its located in. There are two main types of comparison experiments, namely replaced and shuffled experiments.

The replaced cases are to test the concentration or dispersion of individuals' real publication records on the epistemic landscape. The simply replaced sequence is generated by randomly picking papers from the original paper set without put-back, based on the paper count per scientist. The second, a rigorously replaced experiment that incorporates cell information, increases the reserve of the local concentration in original trajectories. We randomly pick cells from the paper set to replace unique cells in the original trajectories. In the modified sequence, the visitation frequency and order of the cell remain the same. The third replaced experiment is designed to check Zipf's law in the visitation frequency of unique cells in one's trajectories. We modify the frequency of unique cells in one's trajectory by random number generation.

In the second scenario, we reshuffle the sequence of individual scientists' publication records to explore potential temporal patterns by comparison. For the same example, the generated sequence after being reshuffled is $(h, f, g, d, a, i, b, c, e)$**.** The other shuffled experiment targets explorative points, not overall track points. As in the example, the exploration points are $(c, d, h)$ and their order is disrupted as $(h, c, d)$. The change of exploration steps before and after the interruption is calculated to check whether exploring new occurs in small steps.

## Supplementary Note 3: Key results under different parameter settings of algorithms

The implementation of this study is mainly based on two algorithms, Doc2vec[10] and UMAP[11]. Here, we intend to check the effect of the parameters of algorithms on our results.

Firstly, we check the dimension of the epistemic landscape. The results in the main text are based on a two-dimensional epistemic landscape, for the convenience of visualization. Here we have performed a 3D dimension reduction with UMAP. Based on the 3D epistemic landscape, we conduct some key experiments to check the robustness of patterns, namely $\Delta r$ distribution, $r_g$ distribution, and the association between $r_g$ and academic performance. As shown in Supplementary Fig. S9, the observations are consistent with those in the main text, suggesting non-sensitivity of the dimension of the epistemic landscape.

Second, the performance of machine learning algorithms often relies on the fine-tuning of parameters. A majority of the parameters are set to the default values in the main text. Doc2Vec has several basic parameters, including the vector dimension, window size, training rounds, etc. Here, we argue that the vector dimension affects the model's ability to learn text content, thus we examine the sensitivity of the results to the vector dimension. The basic UMAP parameters include the number of neighbors, reduced dimension, and distance metric. The parameter of the number of neighbors controls how UMAP balances local versus global structure in the data. The parameter of metric controls how distance is computed in the ambient space of the input data. Thus, we pay close attention to these two parameters. The parameter space is infinite, and here our focus is to verify whether we can still observe consistent results within a suitable and reasonable range of parameter settings. The parameter values are taken as shown in Supplementary Table S4 and one parameter is changed each time.

Supplementary Fig. S10 shows the results under different parameters, from left to right for the vector dimension equal to 200, the number of neighbors equal to 50, and the distance metric being Euclidean, respectively. The adjustment of the parameters (the vector dimension and number of neighbors) merely alters the map's shape, but the knowledge structure is similar to the original. Correspondingly, the results on distributions of $\Delta r$ and $r_g$ and the association with academic performance obtained with both parameters is consistent with the results in the main text. However, learning based on the metric parameter being Euclidean is unable to capture the knowledge structure, resulting in deviated statistical patterns due to the inappropriateness of the Euclidean metric in measuring the difference in values of each dimension of the document vectors.

The above robustness check on algorithm parameter settings verifies that the observed patterns have no serious dependence on the careful fine-tuning of the algorithm parameters.

## Supplementary Note 4: Extended experiments with other four large-scale disciplines

In this paper, we focus on six disciplines, including Computer Science, Physics, Chemistry, Biology, Social Science, and Multidisciplinary Science. Apart from Physics, we first investigate Chemistry, Biology, Social Science, and Multidisciplinary Science with the method of Doc2Vec and UMAP based on semantic proximity, as deployed in CS data. In Supplementary Fig. S11-S14, four epistemic landscapes are annotated by key phrases extracted from the text content of papers with the KeyBERT algorithms.

Extended experiments in these four disciplines include two spatial characteristics ( $\Delta r$ distribution, $r_g$

distribution), the generation of the track-point heatmap by the BEE model, and the association between $r_g$ and academic performance. From Supplementary Fig. S15-S18, it could be seen that the $\Delta r$ distributions are fitted better by an exponential function, and the $r_g$ distributions are fitted by the Gaussian or the lognormal, both without fat-tailed characteristics. These non-Lévy-flight observations are in line with those observed in Computer Science. We also observe a better approaching generated result of our proposed BEE model in four disciplines, which also indicates there exist the frequency bias and the knowledge-proximity bias in scientists' mobility in these disciplines.

On association with academic performance, the results are similar across four disciplines, except for a slight difference in the trend of publication count indicator at greater $r_g$ value. In the four disciplines, with the increase of $r_g$, we generally see the increase of disruption, the decrease of average citation, the non-sustained increase of publication count and h-index. For publication count, we observe a consistent increase until the middle $r_g$ across all four disciplines. But as $r_g$ goes from median to greater, we observe that the indicator remains almost unchanged in Computer Science, decreases slightly in Chemistry and Biology, and continues to increase in Social Science and Multidisciplinary Science. In summary, despite slight differences in these trends, consistent conclusions can be drawn.

## Supplementary Note 5: Extended experiments in Physics data with the graph embedding method

To test whether the experiment results are independent of the methods of construction of the epistemic landscape, we analyze Physics data with a different embedding method. We take advantage of the Physics and Astronomy Classification Scheme (PACS) codes to construct the epistemic landscape of the Physics discipline, which further precludes the dependence of the mobility pattern on the way the map is constructed. PACS codes are used by the American Physical Society (APS) to classify topics in Physics. Each paper is marked with three or more PACS codes, such as 03.65.Aa. Here, we mainly use the first four digits of the PACS code. Then we construct the PACS co-occurrence network and vectorize the nodes with the network embedding method Node2Vec[13]. Finally, the two-dimensional coordinates of each PACS are obtained with UMAP[11]. Then the coordinates of the papers are obtained by averaging the coordinates of all PACS of the papers. Therefore, the trajectories of scientists are traced based on their publication records.

Supplementary Fig. S19a is the constructed epistemic landscape of Physics, where PACS codes are colored according to their first two digits, and the gray shade represents the distribution of paper points. The distribution of PACS codes also forms a clear clustering structure. The epistemic landscape depicts a clear structure of topics ranging from electromagnetism, condensed matter to nuclear physics. Moreover, in Supplementary Fig. S19b, these three trajectories of scientists with different publication counts present similar localized characteristics as shown in the main text. Next, based on the obtained trajectories, we quantify spatial scale features (Supplementary Fig. S20) and temporal patterns (Supplementary Fig. S21). The non-Lévy flight characteristics presented are roughly consistent with the results derived from the other four discipline data. The slight difference is that the $r_g$ distributions are more approximately lognormal than normal and the small $\Delta r$ are more dominant. These disparities are likely due to the variability of the structures of these two domains. Then, we find the mechanisms of the BEE model can generate more approaching density heat maps to the real Physics landscape (Supplementary Fig. S22). The association in Supplementary Fig. S23 is also consistent with those in the main text, despite the large oscillations in the trends due to the small number of scientists analyzed.

In summary, the results show that the mobility patterns and the Doc2Vec+UMAP epistemic-landscape-construction method exhibited in the main text can also be observed in the case of the APS dataset for the Physics discipline and the Node2Vec+UMAP epistemic-landscape-construction method.

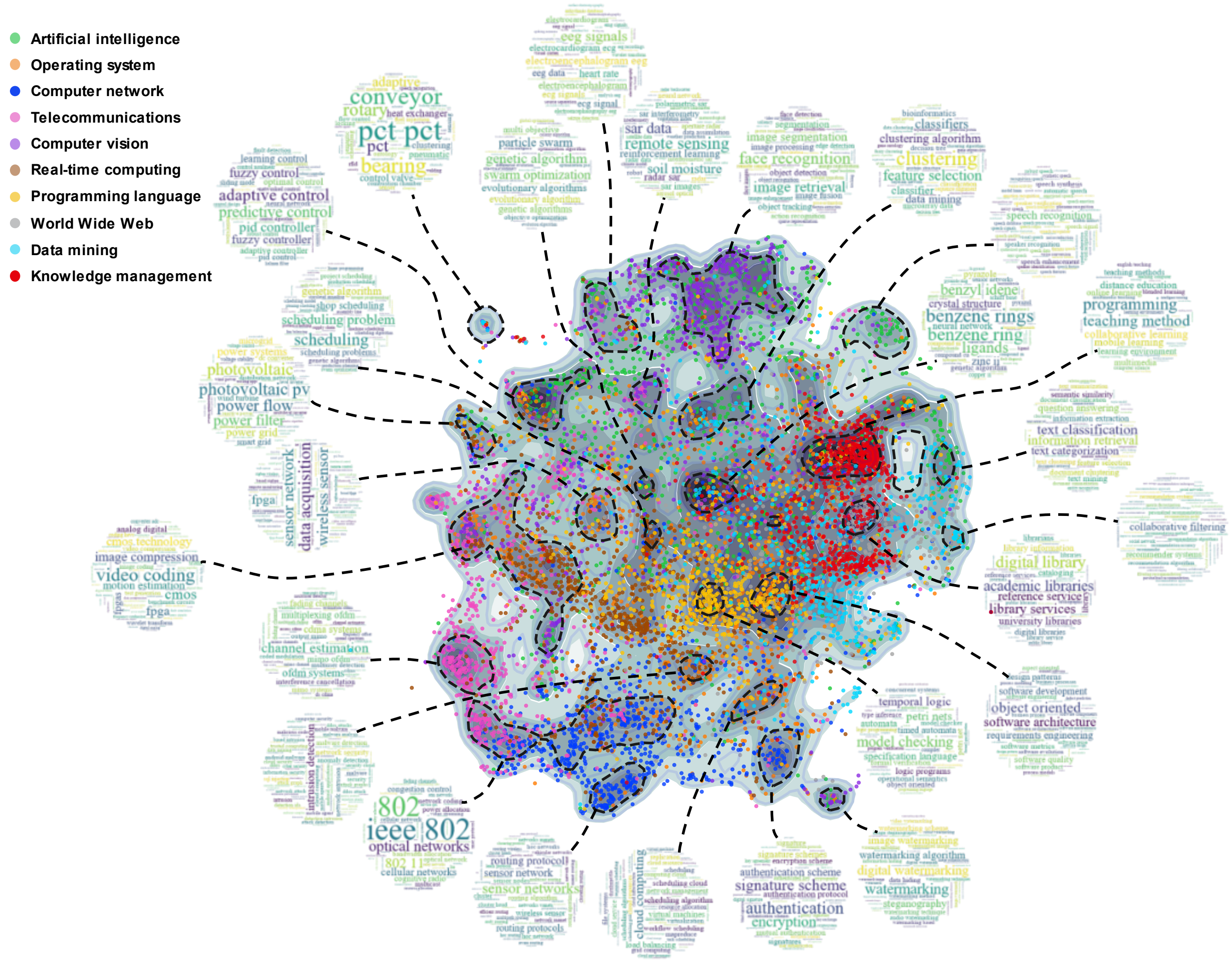

**Supplementary Fig. S1 | The CS epistemic landscape marked by key-phrase clouds.** The bottom is the density distribution of CS paper points, where the darker the color is, the greater the density. Scattered across it are several points sampled from the data set, whose colors indicate their subfields. Here, for clarity, only ten subfields with larger sizes are presented. With the KeyBERT algorithm[2], three key-phrases for each paper are extracted from its title and abstract. Furthermore, several regions with a higher density of paper points are selected and circled on the epistemic landscape. To annotate the research content of the regions on the epistemic landscape, we draw key-phrase cloud maps with the key-phrases of papers in each circled region. In key-phrase clouds, the font size indicates the frequency of the key phrase.

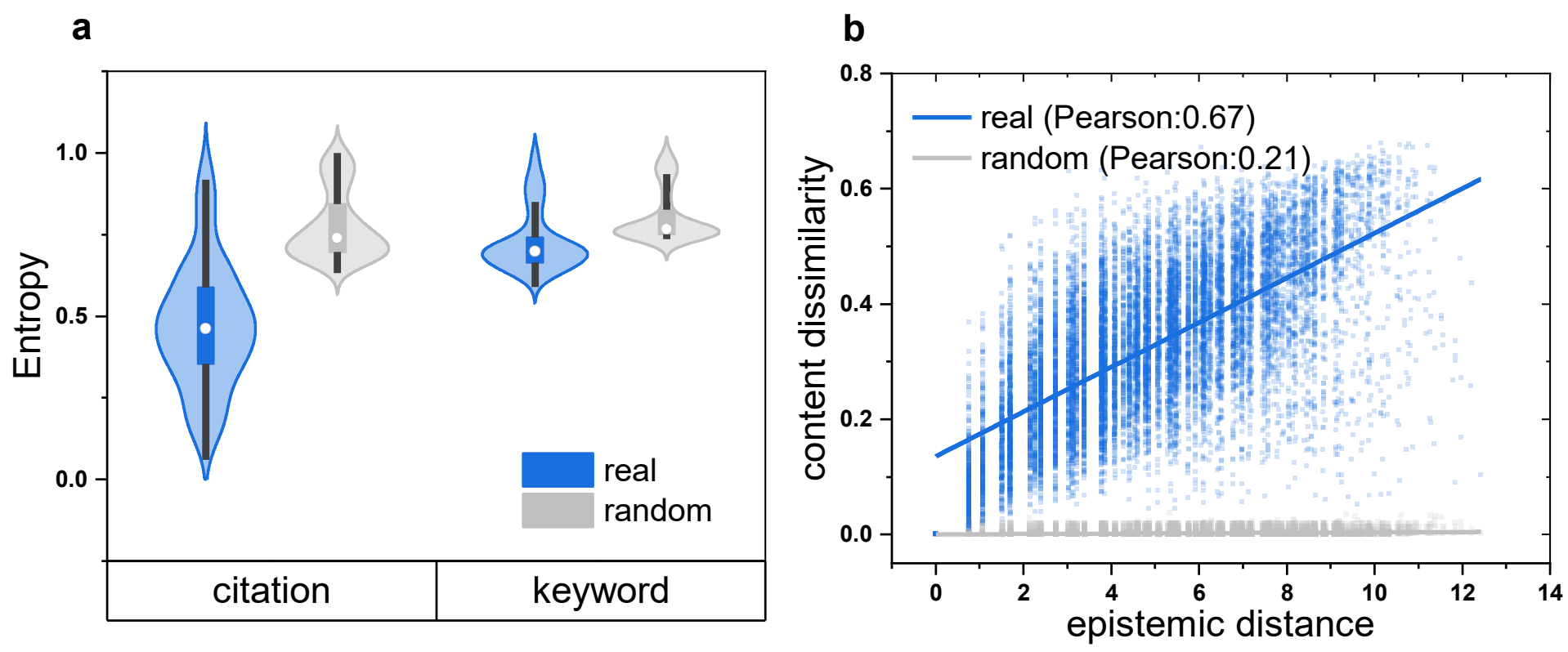

**Supplementary Fig. S2 | The validity of the epistemic landscape of the CS field.** In addition to visually observing the rationality of the knowledge structure shown in the landscape, we systematically quantify the validity of the CS epistemic landscape based on mesh division. **a**, Using the entropy index, it is found that the purity of distribution of FoS labels provided by MAG and citation communities obtained by community detection in the direct citation network in the same cell are both higher than that in the random case, indicating that topics in the same cell share similar research content. **b**, We construct FoS vectors of each cell to represent the research content of the cells with all FoS labels of the papers located in one cell. The cosine distance between each pair of cells is then calculated. It is found that the content distance and epistemic distance between the cell pairs are significantly positively correlated, with a higher correlation coefficient of 0.67 than the random case value of 0.21.

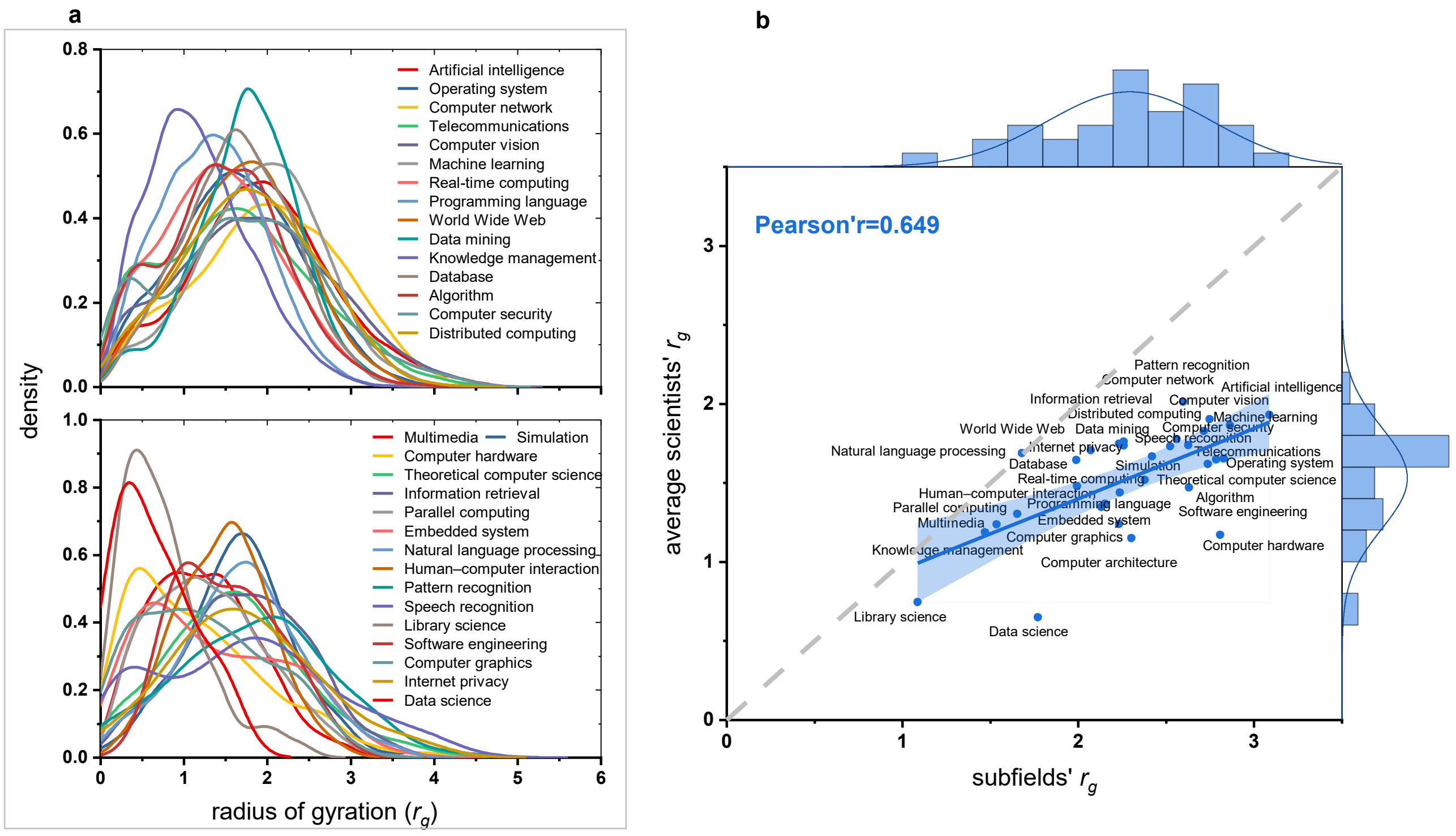

**Supplementary Fig. S3 | The inter-subfield gap in scientists' mobility in CS. a**, The distribution of scientists' $r_g$ in 34 subfields. For each scientist, by aggregating level-1 FoS labels of their publications, we take the label with the highest frequencies as their unique subfield. The distributions of scientists' $r_g$ under each subfield is plotted, ordered by decreasing subfield size. These curves are close to the Gaussian distribution, presenting a different steepness. **b**, the association between subfields' $r_g$ and scientists' $r_g$ in each subfield. For each subfield, the $r_g$ of the distribution of its papers is calculated, and the average value of its scientists' $r_g$ is calculated. Both $r_g$ distributions of subfields present Gaussian characteristics. Each point is located below the diagonal (gray line), except for the natural language processing subfield where the values are located on the diagonal. This result indicates the average values of scientists' $r_g$ are smaller than the papers' $r_g$ for each subfield. These two kinds of $r_g$ are positively correlated, with a Pearson correlation coefficient of 0.649.

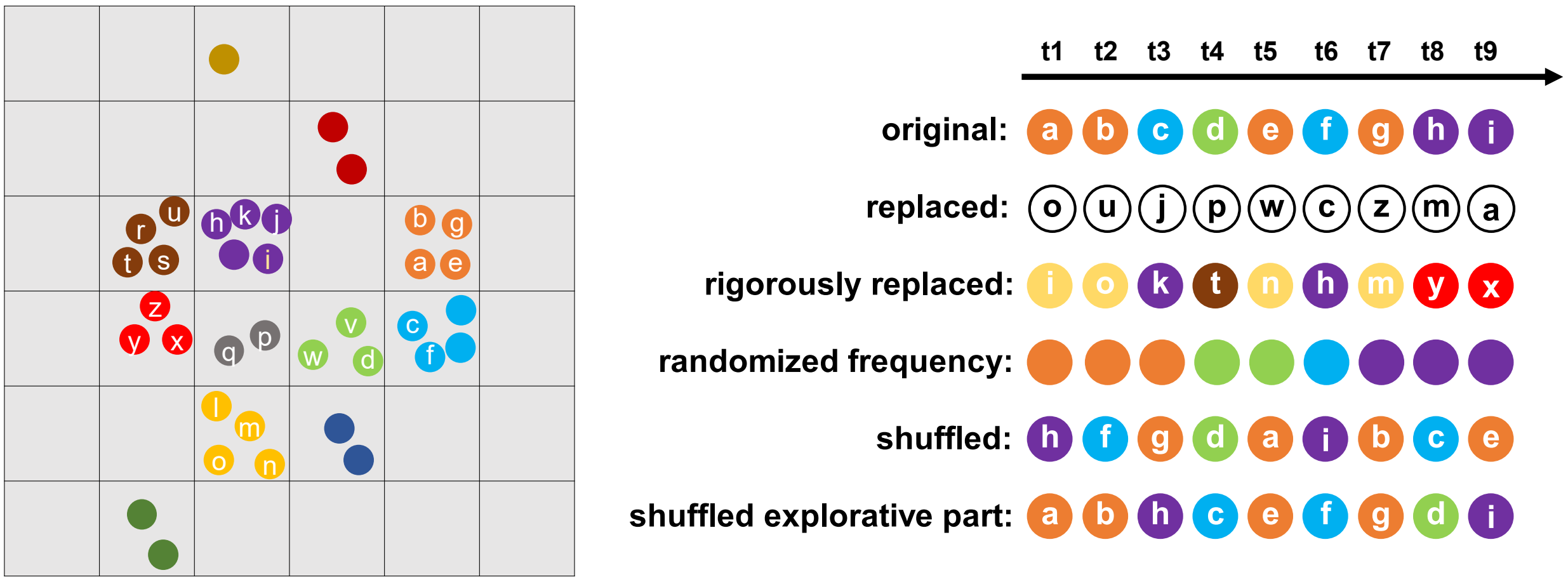

**Supplementary Fig. S4 | An illustration of the creation of five modified publication sequences.** Five kinds of controlled experiments are set, namely replaced case, rigorously replaced case, randomized frequency case, shuffled case, and merely shuffling the order of explorative tracks. The left panel is a diagram of the gridded landscape. The gray represents the epistemic landscape, each circle represents a paper point, and the color of the circle indicates the cell it belongs to. The right panel shows an example of a scientist's trajectory and five modified trajectories.

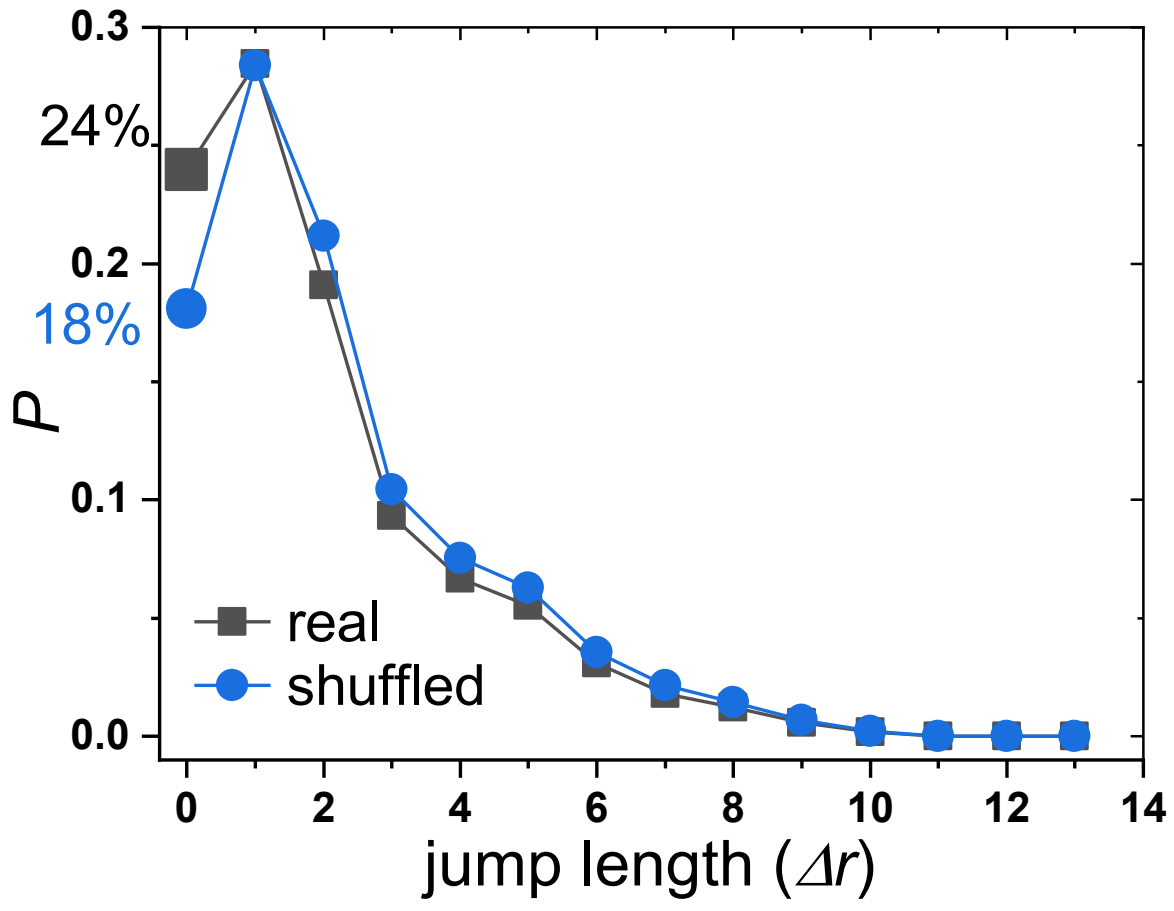

**Supplementary Fig. S5 | The jump-length distributions in real and shuffled cases.** In the real case, the proportion of $\Delta r$ approximately equal to 0 is 24%, higher than 18% in the shuffled case, where the track points in one's trajectory are randomly disordered. This indicates that there is a clear tendency for scientists to publish consecutively on a particular topic.

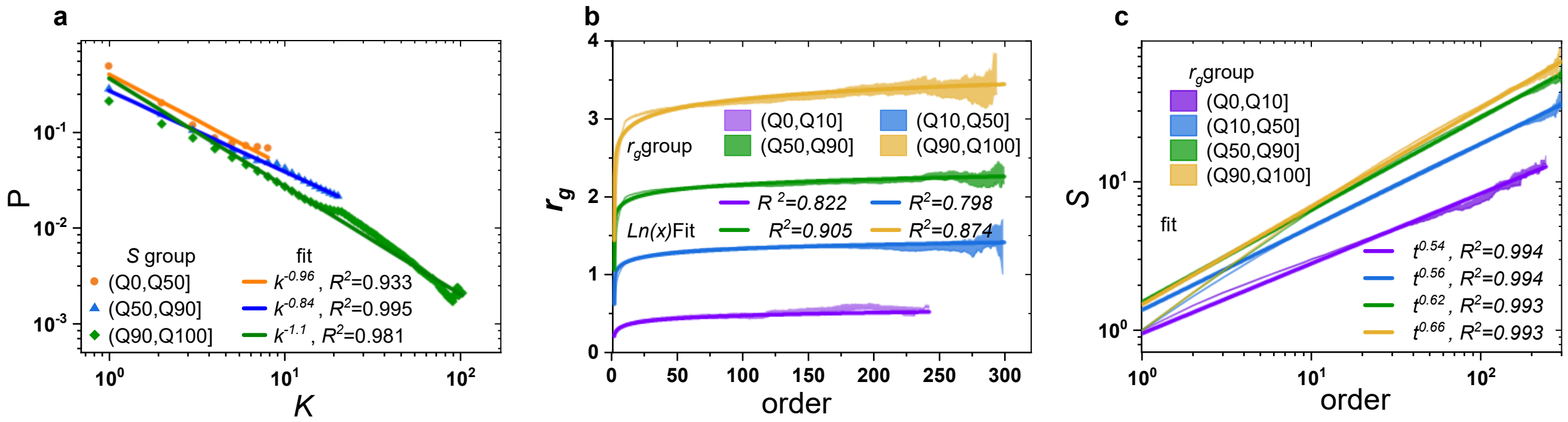

**Supplementary Fig. S6 | The fitting of distributions in subgroups in Fig 3a-c. a**, The visitation frequency $p$ of the $k$th most visited cell in each scientist's trajectory for $S$ groups. $S$ represents the number of unique cells that appear in one's trajectories. We divide scientists into three groups at the $S$ quantile of 0–50, 50–90, and 90–100. The data in each group are well approximated by $k^{-0.96}, (R^2 = 0.933)$, $k^{-0.84}, (R^2 = 0.995)$, and $k^{-1.1}, (R^2 = 0.981)$, respectively. **b**, The growth of $r_g$ as publication aligns with the logarithmic fitting form $A + Bln(t + C)$. These scientists are divided into four groups according to the quantile of their $r_g$. The $R^2$ of fitting for each group are 0.822, 0.798, 0.905, and 0.874. **c** The growth of the $S$ is fitted with $t^\alpha$, respectively $t^{0.54}, (R^2 = 0.994)$, $t^{0.56}, (R^2 = 0.994)$, $t^{0.62}, (R^2 = 0.993)$, and $t^{0.66}, (R^2 = 0.993)$.

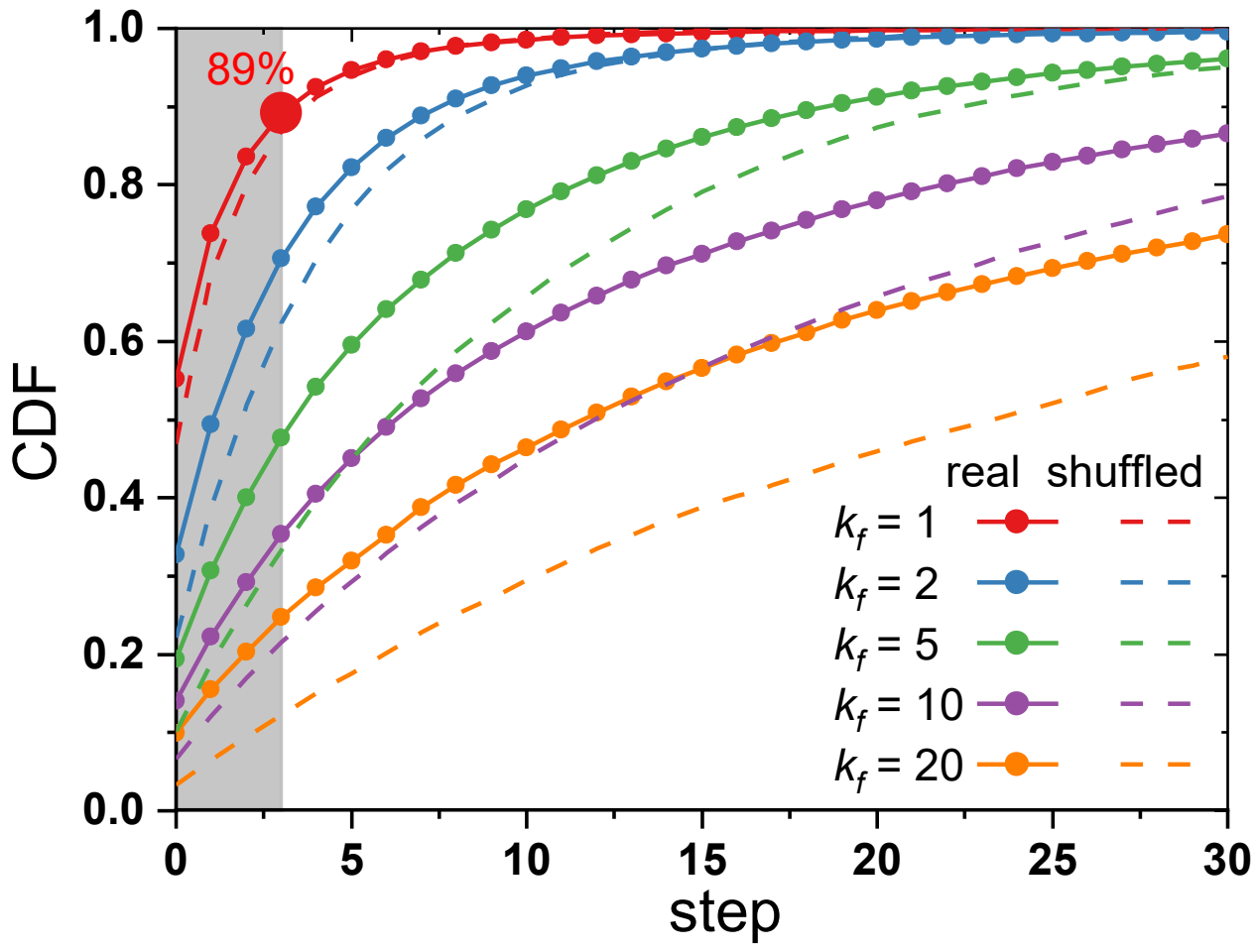

**Supplementary Fig. S7 | Fraction of returns to the $K_f$ most-visited cells occurring after steps.** Another way to see the frequency effect is by analyzing the correlation between the number of moving steps between two visits to a cell. The moving step is 0 if the last visited cell is the same as the one in this step. Regarding high-frequency visited cells, we mainly examine the cells with $K_f = 1, 2, 5, 10, 20$ in one's trajectory. About 89% of the returns to the most frequently visited topics ($K_f = 1$) occurred within 3 steps. We can see scientists' strong tendencies to return to their most-visited cells within very few steps, which is more evident than in the shuffled case.

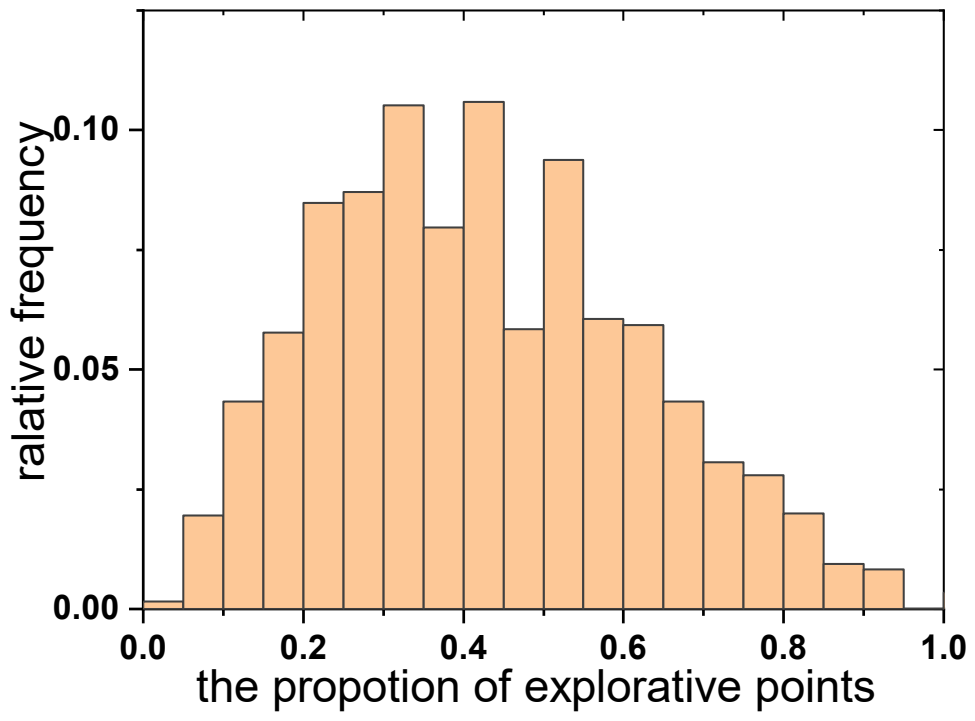

**Supplementary Fig. S8 | Percentage of exploration trajectories in scientists' trajectories.** We calculate the fraction of explorative points in one's trajectory. In simulating one's trajectory, we use this fraction to determine the type of next moving step.

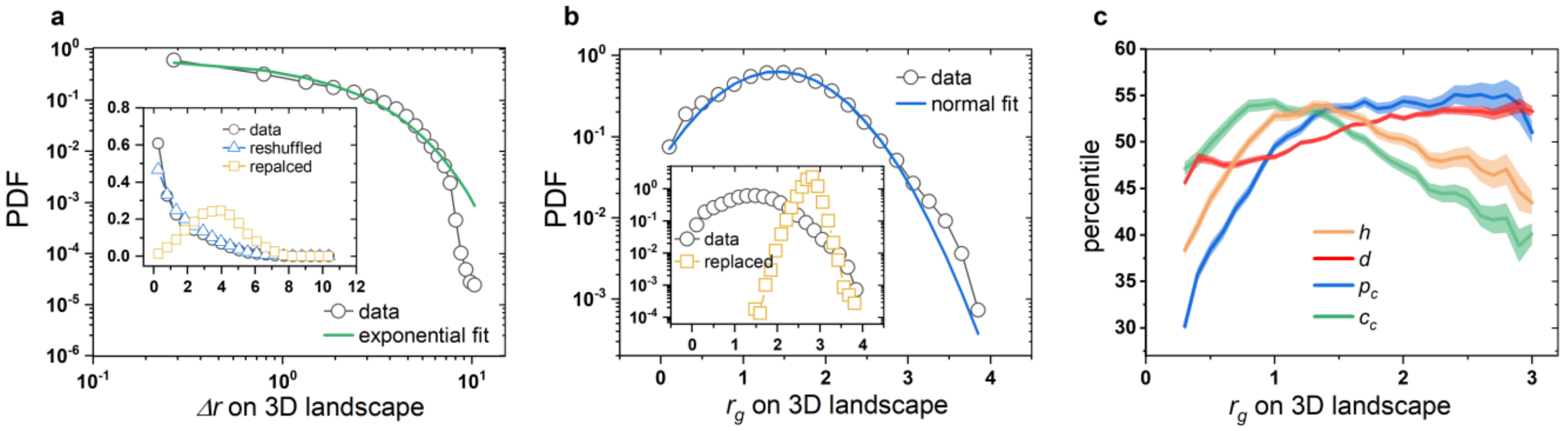

**Supplementary Fig. S9 | The consistent pattern of scientists' mobility is observed in the three-dimensional CS epistemic landscape. a**, The $\Delta r$ still shows an exponential distribution, with a high proportion of small-size jumps absent in controlled experiments. **b**, The $r_g$ distribution is approximate to normal, which disappears in controlled experiments. **c**. The $h$ and average $c_c$ both show first increasing and then decreasing trends, the $p_c$ shows an increasing trend with saturation, and the $d$ shows an increasing trend, all of which correspond to that in the main text. In a word, we observe the same pattern as that in the two-dimensional epistemic landscape.

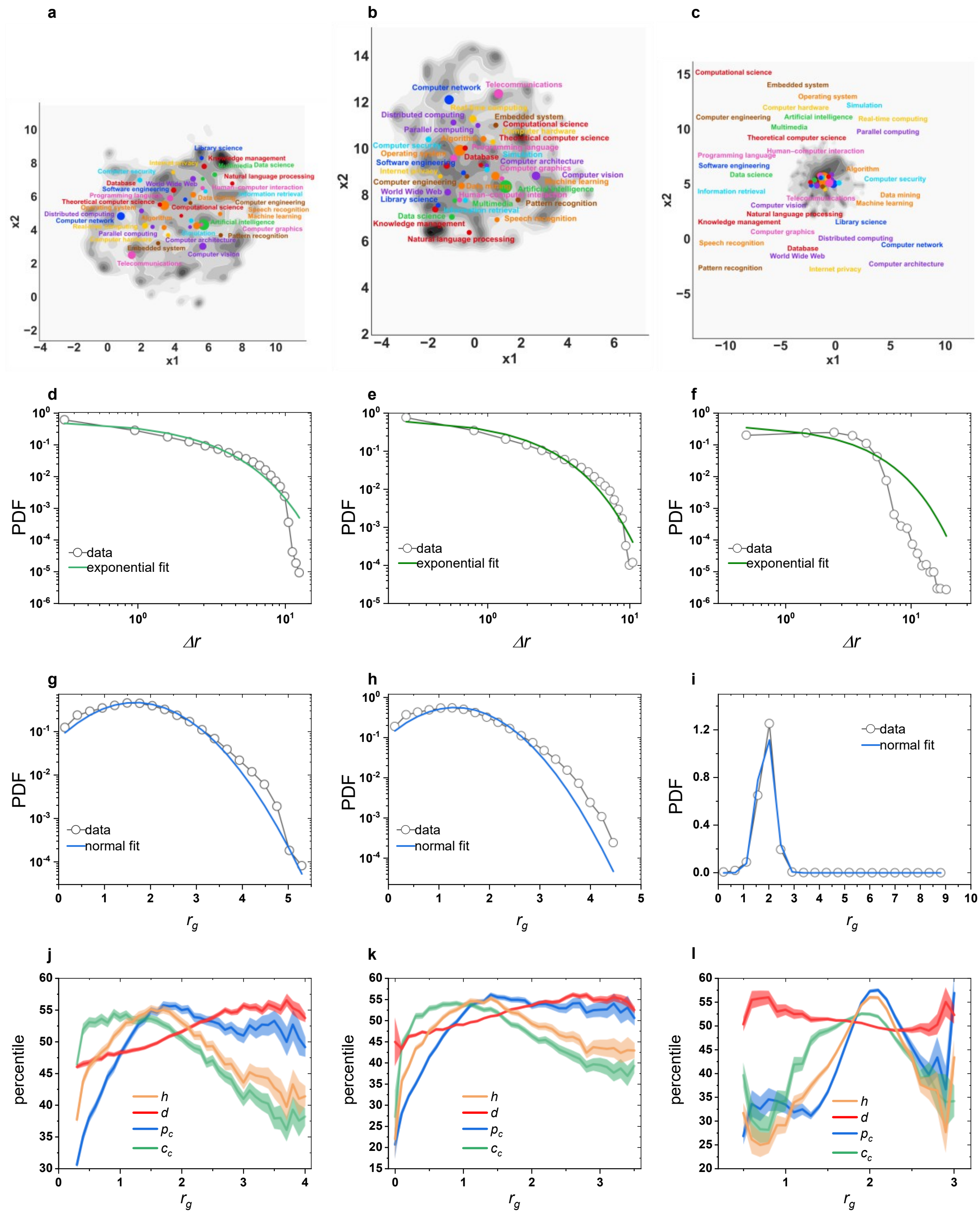

**Supplementary Fig. S10 | The key results under three algorithm parameter settings in CS data.** Based on the original parameter settings, we change one of the parameters at a time to test the pattern's robustness. There are three parameters of concern here, namely the vector dimension parameter of Doc2Vec, number of neighborhoods parameter, and distance metric parameter of UMAP. The original settings of these three parameters are 100, 15, and cosine distance, respectively. The consistent patterns are observed when the vector dimension is set at 200 (**a**, **d**, **g**, **j**) and the number of neighborhoods is set at 50 (**b**, **e**, **h**, **k**). The results are inconsistent at the metric parameter of the Euclidean distance (**c**, **f**, **i**, **l**). In short, the observed patterns are not a fluke in particular parameter settings.

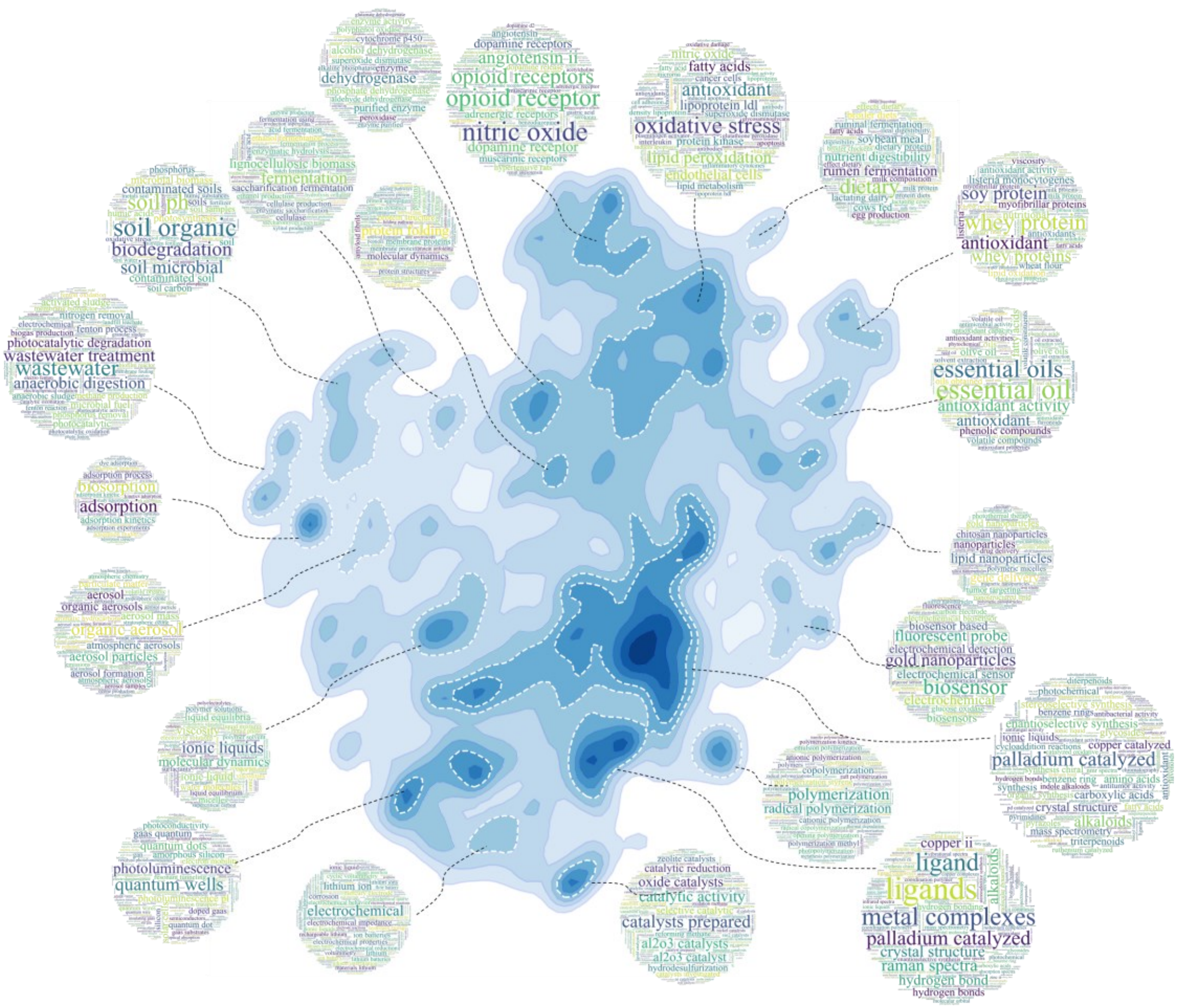

**Supplementary Fig. S11 | The Chemistry epistemic landscape marked by key-phrase clouds.** The bottom is the density distribution of Chemistry paper points, where the darker the color is, the greater the density. With the KeyBERT algorithm[2], three key-phrases for each paper are extracted from its title and abstract. Furthermore, several regions with a higher density of paper points are selected and circled on the epistemic landscape. To annotate the research content of the regions on the epistemic landscape, we draw key-phrase cloud maps with the key-phrases of papers in each circled region. In key-phrase clouds, the font size indicates the frequency of the key phrase. The epistemic landscape depicts an aggregated structure of topics. The center is located by the research topics related to "palladium-catalyzed reactions", "crystal structure" and "ligand". The topics located at the periphery range from "biosensor", and "antioxidant" to "quantum wells".

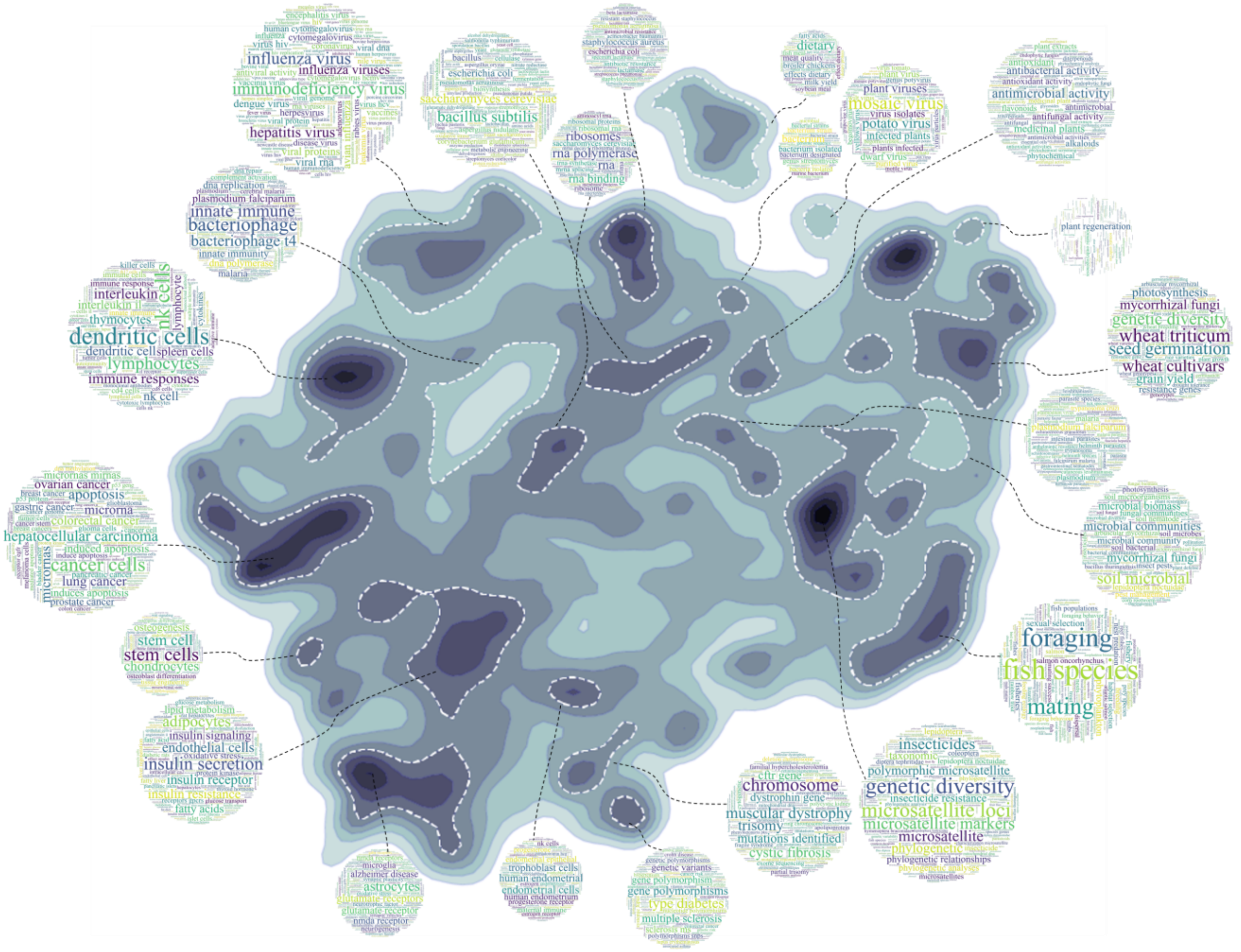

**Supplementary Fig. S12 | The Biology epistemic landscape marked by key-phrase clouds.** The bottom is the density distribution of Biology paper points, where the darker the color is, the greater the density. With the KeyBERT algorithm2, three key-phrases for each paper are extracted from its title and abstract. Furthermore, several regions with a higher density of paper points are selected and circled on the epistemic landscape. To annotate the research content of the regions on the epistemic landscape, we draw key-phrase cloud maps with the key-phrases of papers in each circled region. In key-phrase clouds, the font size indicates the frequency of the key phrase. The center is located by the research topics related to “genetic diversity”, “RNA polymerase” and “cancer cells”. The topics located at the periphery are ranging from “foraging behavior”, “influenza virus” and “Alzheimer disease” to “disease of immune system”.

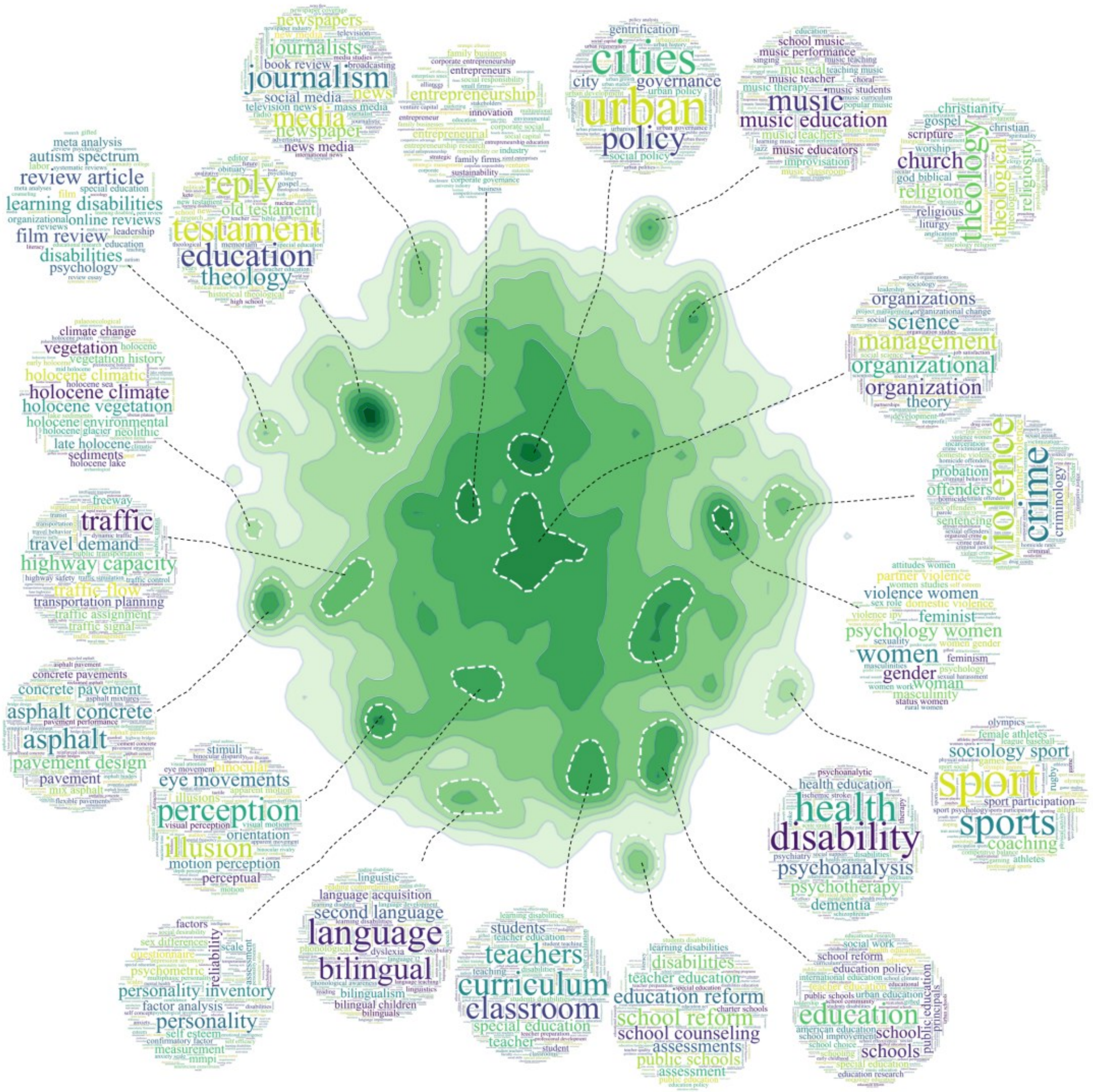

**Supplementary Fig. S13 | The Social Science epistemic landscape marked by key-phrase clouds.** The bottom is the density distribution of Social Science paper points, where the darker the color is, the greater the density. With the KeyBERT algorithm2, three key-phrases for each paper are extracted from its title and abstract. Furthermore, several regions with a higher density of paper points are selected and circled on the epistemic landscape. To annotate the research content of the regions on the epistemic landscape, we draw key-phrase cloud maps with the key-phrases of papers in each circled region. In key-phrase clouds, the font size indicates the frequency of the key phrase. The center is located by the research topics related to "organization management", "urban governance", "health and psychoanalysis theory", "social psychological and personality science" and "entrepreneurship". The topics located at the periphery range from "education reform", and "language study" to "theology and religious study".

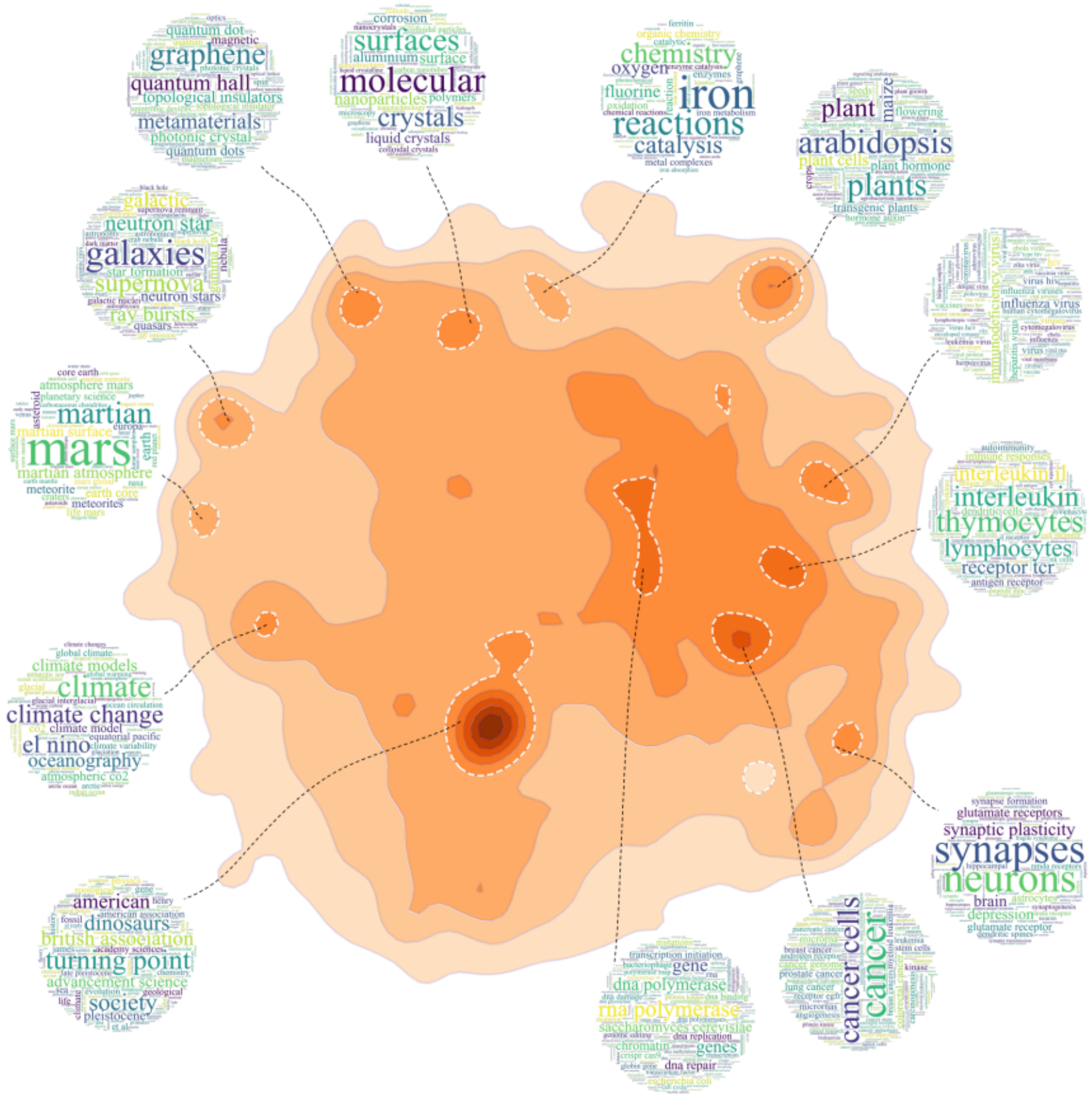

**Supplementary Fig. S14 | The Multidisciplinary Science epistemic landscape marked by key-phrase clouds.** The bottom is the density distribution of Multidisciplinary Science paper points, where the darker the color is, the greater the density. Scattered across it are several points sampled from the data set, whose colors indicate their subfields. Here, for clarity, only ten subfields with larger sizes are presented. With the KeyBERT algorithm[2], three key-phrases for each paper are extracted from its title and abstract. Furthermore, several regions with a higher density of paper points are selected and circled on the epistemic landscape. To annotate the research content of the regions on the epistemic landscape, we draw key-phrase cloud maps with the key-phrases of papers in each circled region. In key-phrase clouds, the font size indicates the frequency of the key phrase. Due to the diversity of research topics in multidisciplinary journals, the topical clusters are shown to be far apart from each other and distributed at the periphery of the whole landscape.

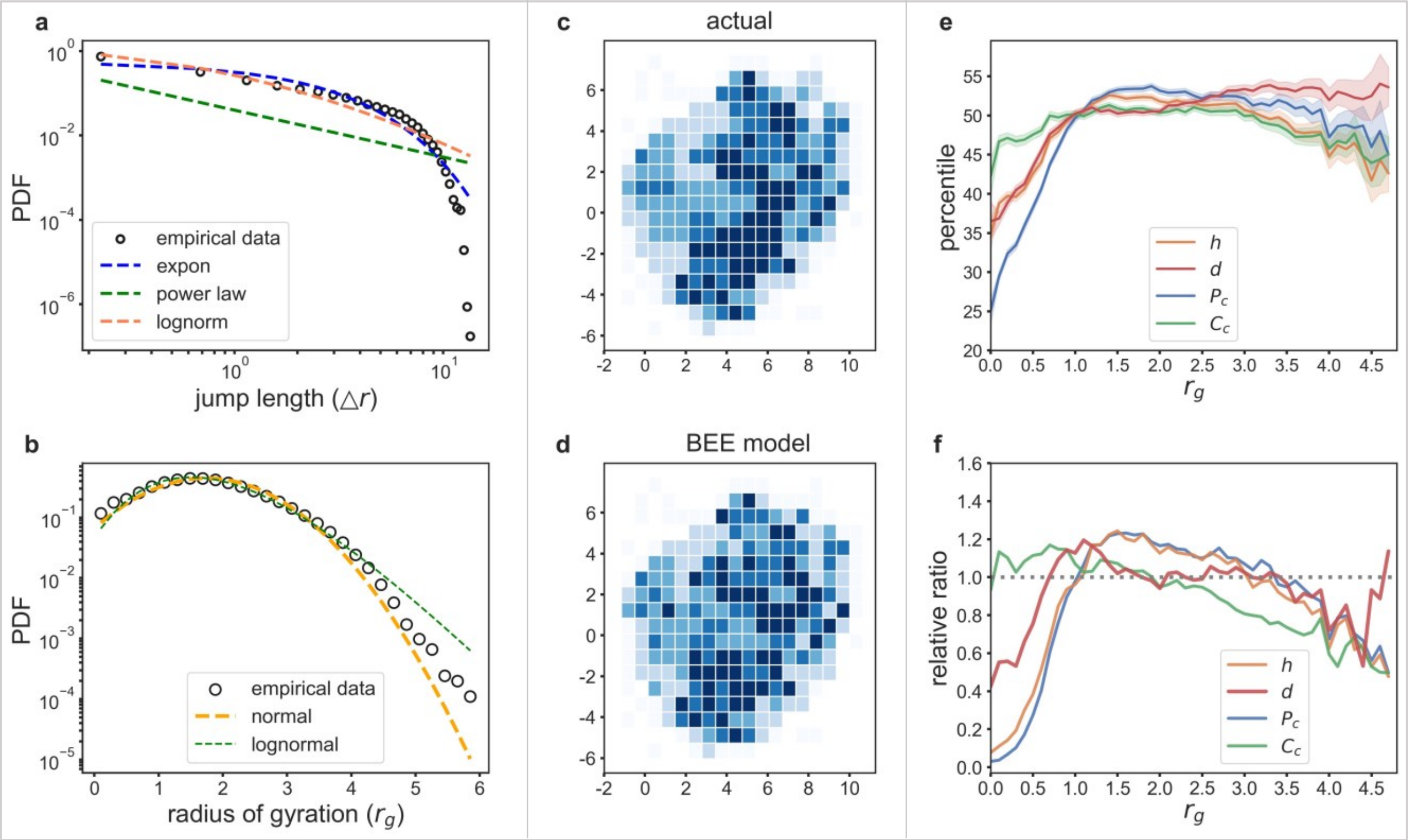

**Supplementary Fig. S15 | The key results in Chemistry data. a**, The $\Delta r$ shows an exponential distribution, rather than power law and truncated power law distribution. **b**, The $r_g$ distribution is approximate to normal, without fat-tail characteristics. **c-d**, The density heat maps of scientists' track points generated by the actual scientists' trajectories and generated by the BEE model. It could be seen the distribution of hotspots on the grid produced by the BEE model is also closer to the original. **e-f**, Correlations between the radius of gyration and four key bibliometric indicators that is the total count of published papers ($p_c$), the average count of citations per paper ($c_c$), h-index ($h$) and mean disruptive value ($d$). **e** is for average percentile case and **f** is for extreme cases (top 10% of scientists with the most bibliometric indicators). With the increase of $r_g$, we observe a decrease of $c_c$, except for the reverse increasing trend when $r_g$ is smaller than 0.2. As $r_g$ increases, $h$ and $p_c$, first increase and then decrease, and $d$ continues to increase. But as the $r_g$ becomes larger, the increase of $d$ gradually slows down.

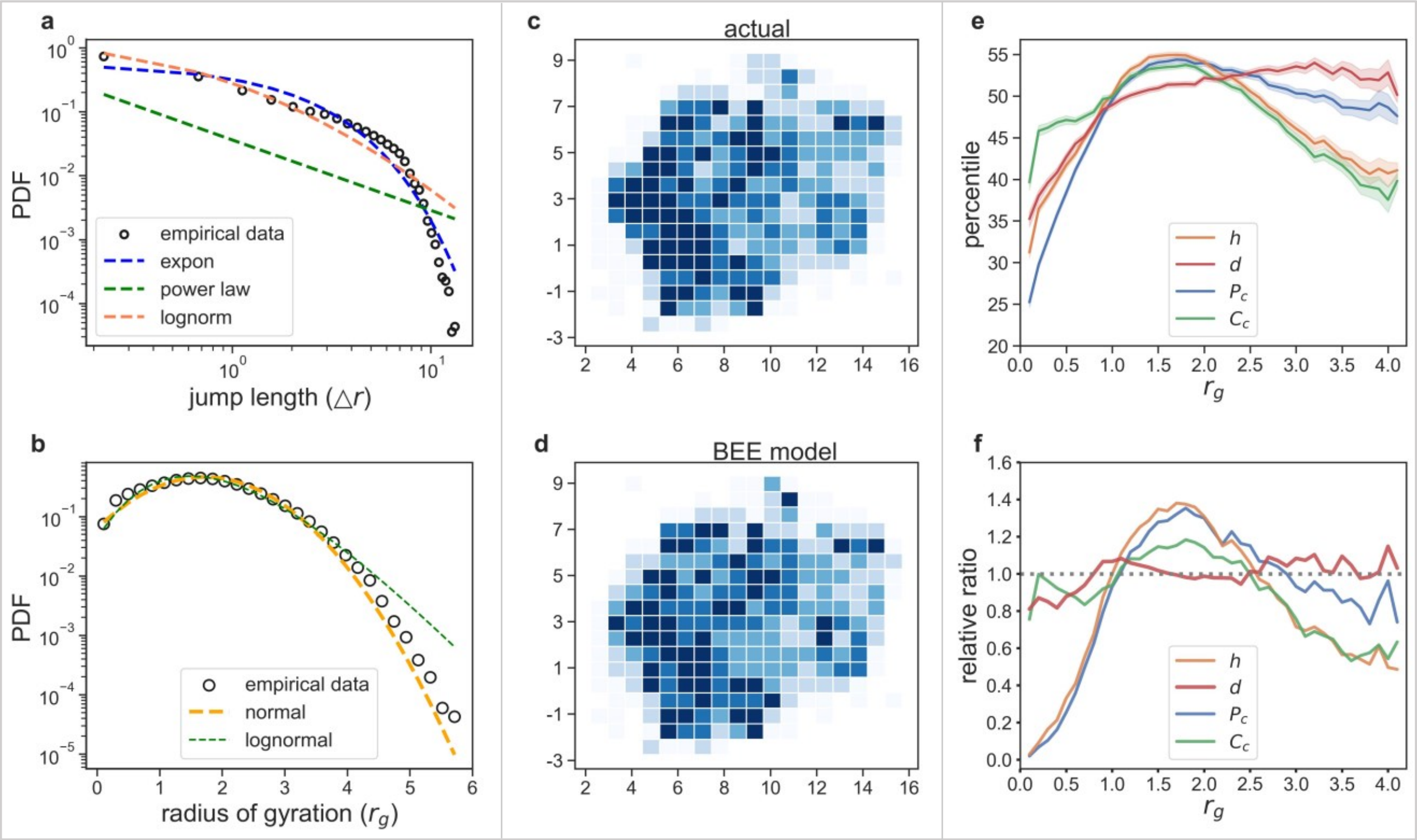

**Supplementary Fig. S16 | The key results in Biology data. a-b**, The spatial characteristics of trajectories are examined, namely jump length ($\Delta r$) and radius of gyration ($r_g$). **a**, The $\Delta r$ shows an exponential distribution, rather than power law and truncated power law distribution. **b**, The $r_g$ distribution is approximate to normal, without fat-tail characteristics. **c-d**, The density heat maps of scientists' track points generated by the actual scientists' trajectories and generated by the BEE model. It could be seen that the distribution of hotspots on the grid produced by the BEE model is also closer to the original. **e-f**, Correlations between the radius of gyration and four key bibliometric indicators that is the total count of published papers ($p_c$), the average count of citations per paper ($c_c$), h-index ($h$) and mean disruptive value ($d$). **e** is for average percentile case and **f** is for extreme case (top 10% of scientists with the most bibliometric indicators). As $r_g$ increases, the $h$ and $c_c$ both show first increasing and then decreasing trends, and $p_c$ first increases sharply then slightly decreases. With the increase of $r_g$, $d$ increases. But as the $r_g$ becomes larger, the increase of $d$ gradually slows down.

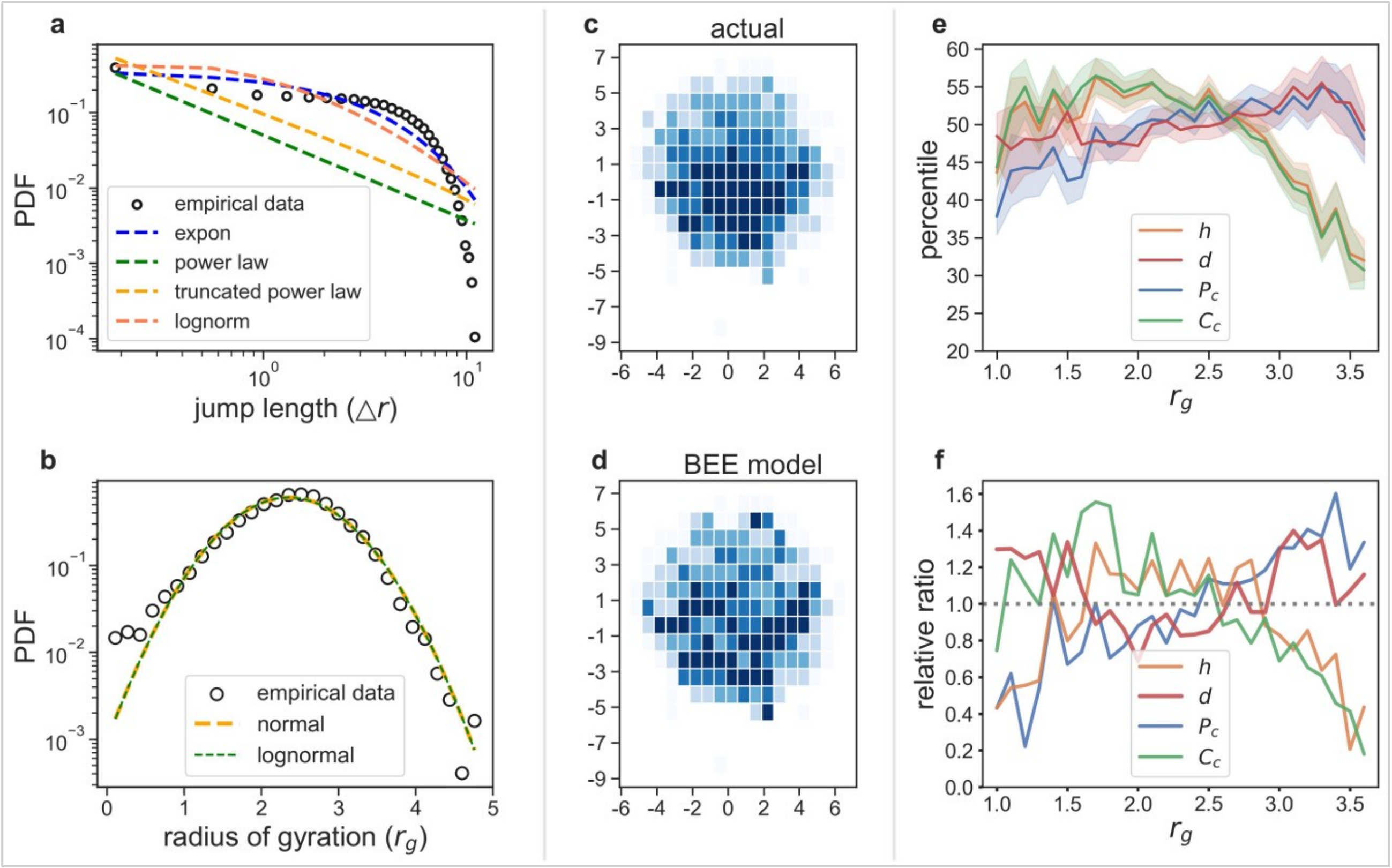

**Supplementary Fig. S17 | The key results in Social Science data. a-b**, Two spatial characteristics of trajectories are examined, namely jump length (Δr) and radius of gyration ($r_g$). **a**, The $\Delta r$ is more approximate to an exponential distribution, than power law and truncated power law distribution. **b**, The $r_g$ distribution is fitted well by lognormal and normal, without fat-tail characteristics. **c-d**, The density heat maps of scientists' track points generated by the actual scientists' trajectories and generated by the BEE model. It could be seen that the distribution of hotspots on the grid produced by the BEE model is also closer to the original. **e-f**, Correlations between the radius of gyration and four key bibliometric indicators that is the total count of published papers ($p_c$), the average count of citations per paper ($c_c$), h-index ($h$) and mean disruptive value ($d$). **e** is for average percentile case and **f** is for extreme case (top 10% of scientists with the most bibliometric indicators). With the increase of $r_g$, it is observed a decrease of $c_c$ and $h$, except for the reverse increasing trend at the very left end of the $r_g$ coordinate. With the increase of $r_g$, we see steady increases of $d$ and $p_c$.

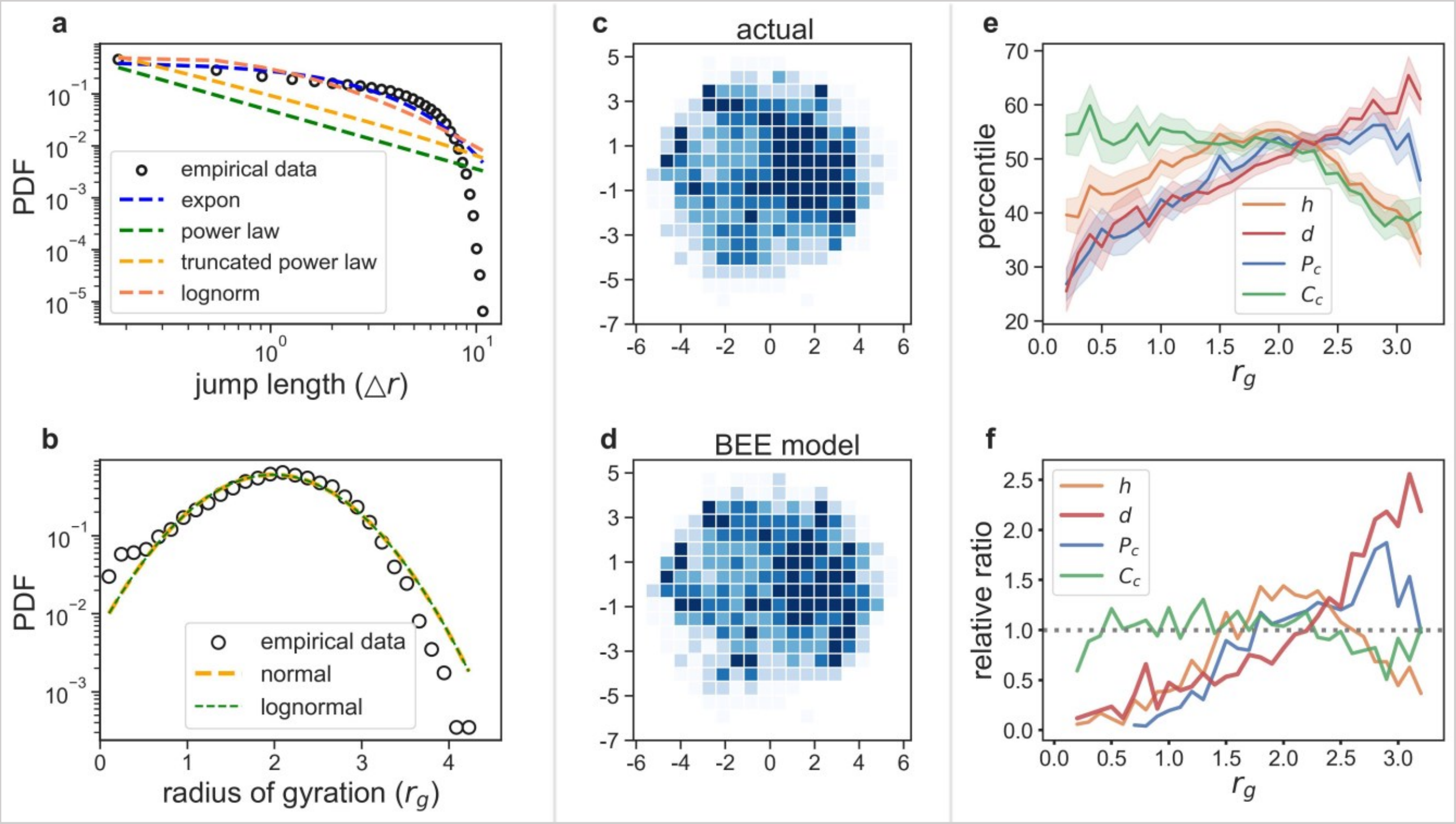

**Supplementary Fig. S18 | The key results in Multidisciplinary Science data. a-b**, Two spatial characteristics of trajectories are quantified, namely jump length (Δr) and radius of gyration ($r_g$). **a**, The $\Delta r$ is approximate to exponential distribution better, than power law and truncated power law distribution. **b**, The $r_g$ distribution is both fitted well by lognormal and normal, without presenting fat-tail characteristics. **c-d**, The density heat maps of scientists' track points generated by the actual scientists' trajectories and generated by the BEE model. It could be seen that the BEE model could reproduce the distribution of hotspots in Multidisciplinary Science data. **e-f**, Correlations between the radius of gyration and four key bibliometric indicators that is the total count of published papers ($p_c$), average count of citations per paper ($c_c$), h-index ($h$) and mean disruptive value ($d$). **e** is for average percentile case and **f** is for extreme case (top 10% of scientists with the most bibliometric indicators). With the increase of $r_g$, $c_c$ decreases, $d$ and $p_c$ increase, and $h$ presents a trend of first increase and then decrease.

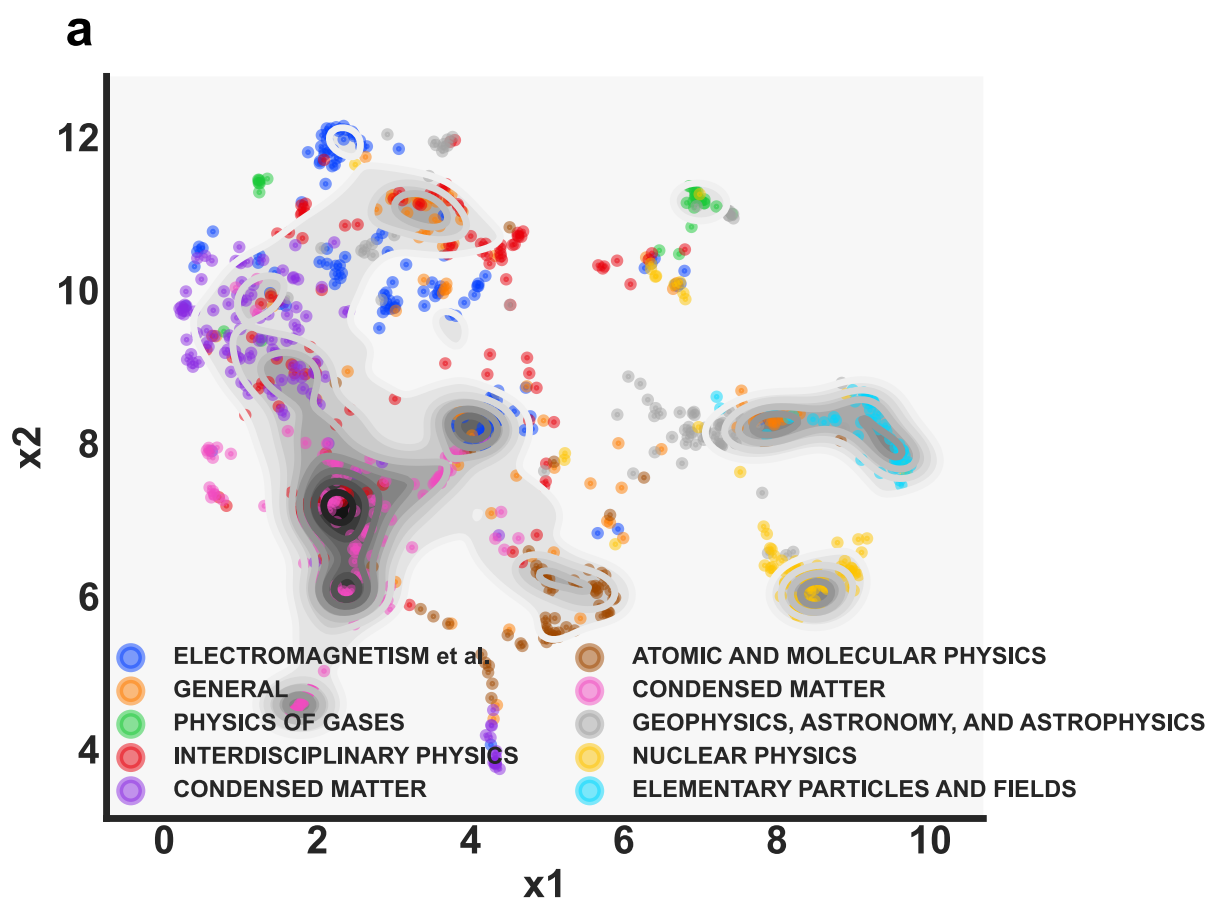

**Supplementary Fig. S19 | The epistemic landscape and scientists' trajectories in APS data. a**, The epistemic landscape is constructed based on the PACS codes co-occurrence network with Node2Vec and UMAP. The dots represent PACS codes colored by the first two digits of PACS codes. The gray shadow represents the distribution of paper points whose coordinates are averaged from the coordinates of its multi-PACS codes. It clearly shows the structure of research subfields ranging from electromagnetism and condensed matter to nuclear Physics. **b**, Three examples of trajectories with different point counts show mobility is merely concentrated in one specific region.

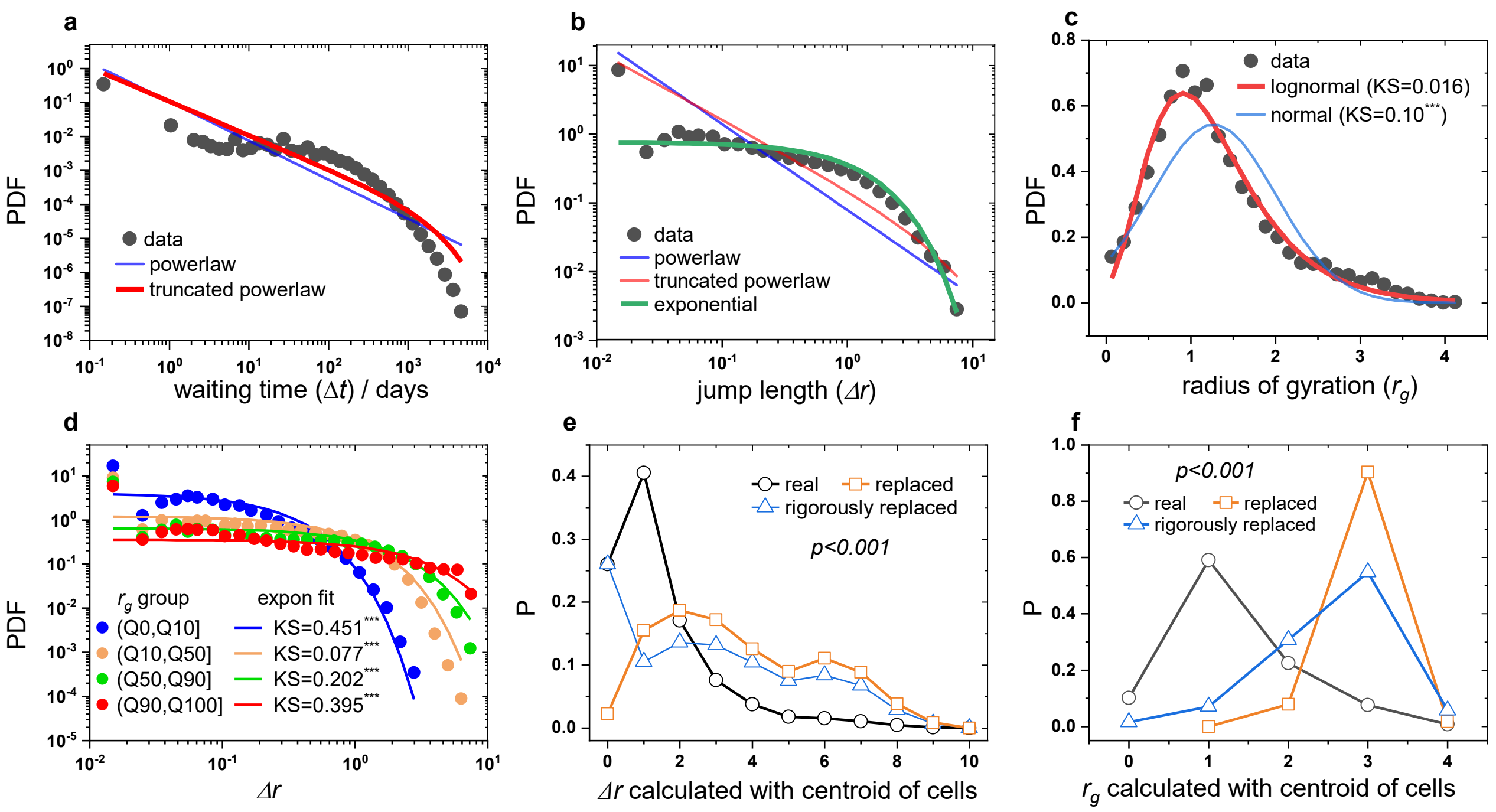

**Supplementary Fig. S20 | The Non-Lévy-Flight characteristics of scientists' mobile trajectories in APS data. a**, The publishing intervals are approximate to a truncated power distribution**. b**, The jump lengths ($\Delta r$) are approximate to an exponential distribution, with a dominant proportion of extremely short jumps. **c.** The distribution of the $r_g$ is much closer to lognormal than normal. **d**, The $\Delta r$ of scientists with different $r_g$ are all exponential mixtures of long and short lengths. **e-f**, Comparisons with two controlled experiments further demonstrate the relative spatial concentration of movement. **a-f**, Again, these observations are consistent with the corresponding result through the CS data.

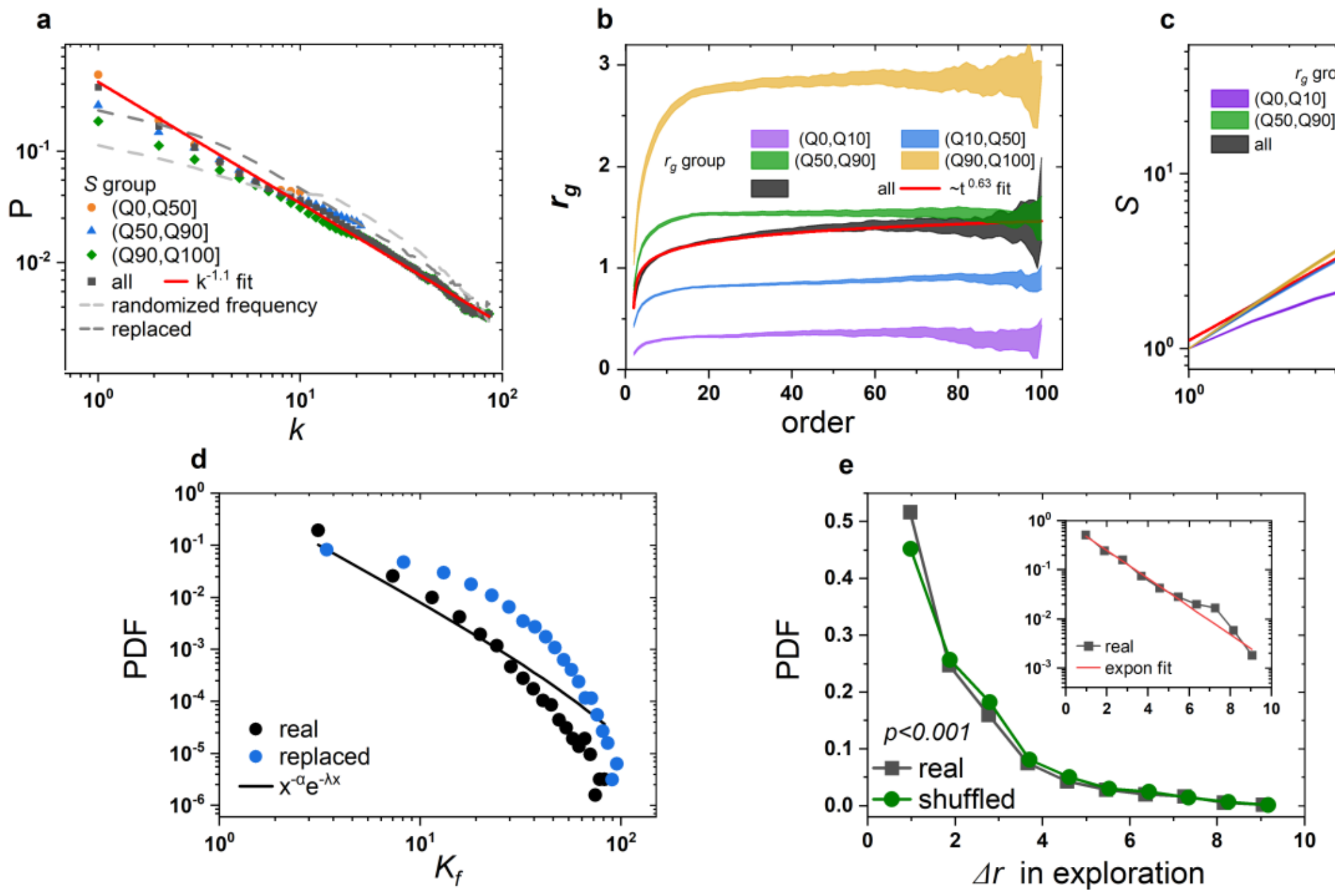

**Supplementary Fig. S21 | The preferences in the mixed exploitation-exploration process in APS data. a**, The visitation frequency of cells in one's trajectory follows Zipf's law. **b-c**, The growth of the number of cells and $r_g$ both present sublinear diffusive processes. **d-e**, We split one's trajectory into the exploration part and exploitation part. If the cell of a given point appears in the trajectory for the first time, we define the point as an exploration point, and vice versa as a deep plowing point. The power law distributed frequency rank ($K_f$) and exponential $\Delta r$ indicate that scientists have frequency bias in exploitation and knowledge proximity in exploration.

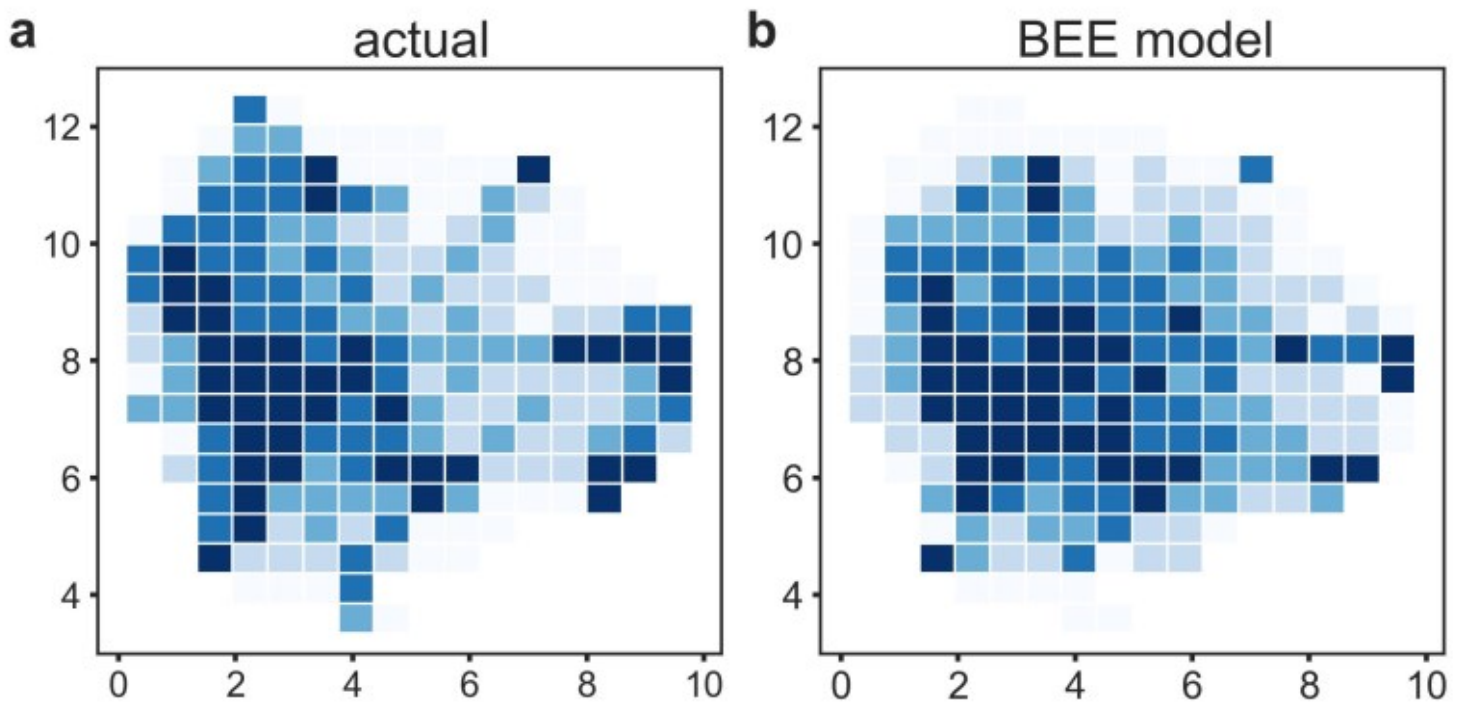

**Supplementary Fig. S22 | The density heat maps of scientists' track points in APS data. a**, generated by the actual scientists' trajectories. **b**, generated by the BEE model. The trajectories generated by our BEE model reproduce the hotspot distribution of the Physics epistemic landscape.

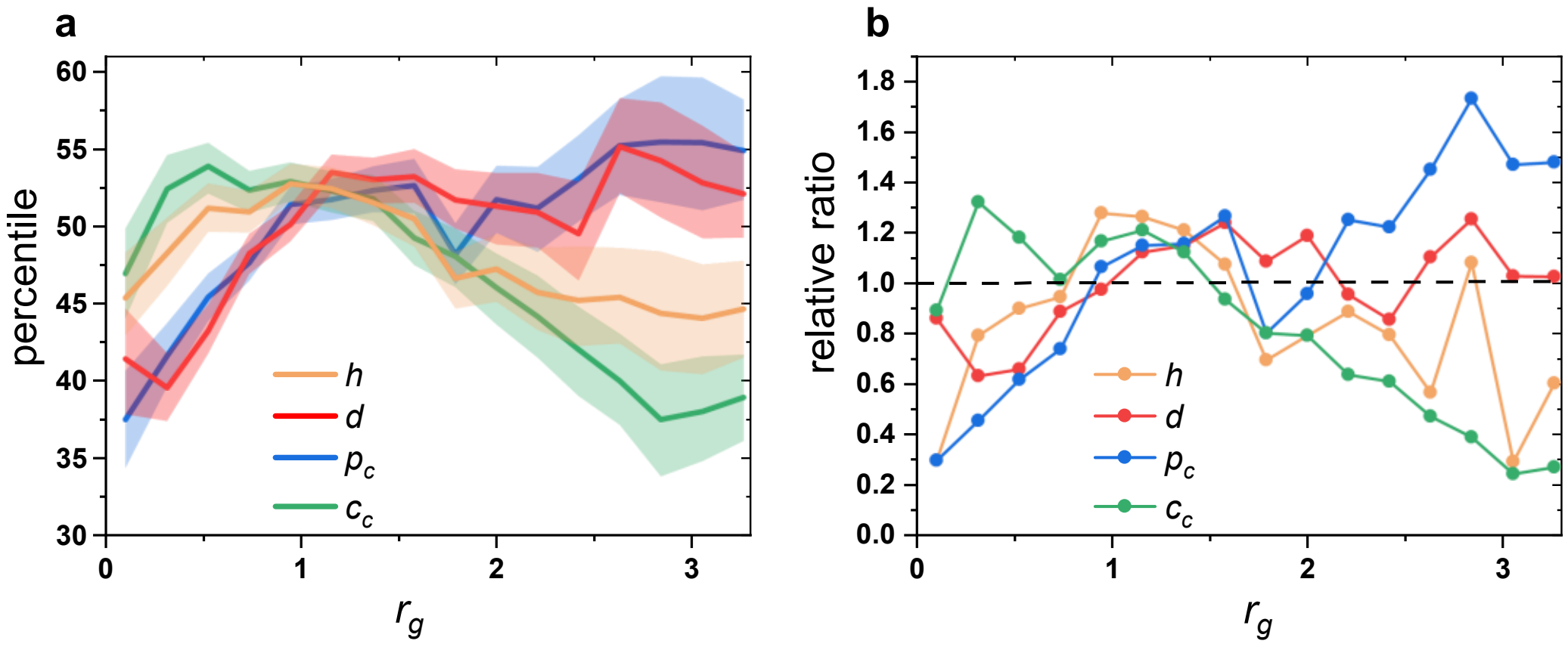

**Supplementary Fig. S23 | The correlation between activity radius and academic performance in APS data.** The academic indexes include paper counts ($p_c$), average citations ($c_c$), h-index ($h$), and the disruptive score ($d$). We calculate the average academic performance of scientists with different $r_g$ (**a**), as well as the percentage of the top 10% of scientists distributed in different $r_g$ groups (**b**). As $r_g$ increase, average $d$ and $p_c$ increase, whereas the average $h$ first increase and then decrease. We observe a decrease of $c_c$ with the increase of $r_g$, except for the reverse increasing trend at the very left end of the $r_g$ coordinate. These trends are similar to those of other five areas, except for the relatively larger oscillations. However, $p_c$ remain unchanged after increasing as $r_g$ increase for other disciplines, but it keeps slow increasing in Physics.

**Supplementary Table S1 | The comparisons of candidate distributions and statistics.** The fitting of candidate heavy-tailed distributions is performed by the method of Clauset et al[6] and the method of Alstott et al[7]. We test the power-law, exponential, and truncated power-law characteristics of waiting time and jump lengths. The goodness of these distribution fits is compared with the "powerlaw" package developed by Alstott et al[7]. The comparison results have two parts. The first is $R$, the log-likelihood ratio between the two candidate distributions. This negative number represents that the data are more likely in the second distribution. The second is the significance value $p$. In the table, $p = 0.0$ due to the $p$-calculated approaches to zero. For $r_g$ distribution, the heavy-tailed fitting failed with the "powerlaw" package. Here we mainly compare normal and lognormal characteristics with the "Disfit" package[8] and "Fitter" package[9]. The goodness of the two distribution fits is measured by RSS and square error.

| **Powerlaw Package** | | | | |
|---|---|---|---|---|
| **indicator** | | **Comparison of Distribution** | **Statistics $(R, p)$** | **Better fit** |
| Waiting time | | ('power law', 'truncated power law') | (-1440.25, 0.0) | truncated power law |
| Waiting time | | ('exponential', ''truncated power law'') | (-717.66, 0.0) | truncated power law |
| jump length | | ('power law', 'exponential') | (-3937.71, 0.0) | exponential |
| jump length | | ('truncated power law, 'exponential') | (-2385.66, 0.0) | exponential |
| jump length | **$r_g$ group** | | | |
| | (Q0,Q10] | ('power law', 'exponential') | (-1224.46, 0.0) | exponential |
| | (Q0,Q10] | ('truncated power law, 'exponential') | (-802.03, 0.0) | exponential |
| | (Q10,Q50] | ('power law', 'exponential') | (-2702.86, 0.0) | exponential |
| | (Q10,Q50] | ('truncated power law, 'exponential') | (-1717.70, 0.0) | exponential |
| | (Q50,Q90] | ('power law', 'exponential') | (-2609.81, 0.0) | exponential |
| | (Q50,Q90] | ('truncated power law, 'exponential') | (-1676.10, 0.0) | exponential |
| | (Q90, Q100] | ('power law', 'exponential') | (-984.06, 0.0) | exponential |
| | (Q90, Q100] | ('truncated power law, 'exponential') | (-609.84, 0.0) | exponential |
| **Disfit package** | | | | |
| **indicator** | | **candidate distributions** | **Statistics (RSS)** | **Better fit** |
| radius of gyration ($r_g$) | | norm | 0.012176 | normal |
| | | lognormal | 0.0180591 | |
| **Fitter Package** | | | | |
| **indicator** | | **candidate distributions** | **Statistics (square error)** | **Better fit** |
| radius of gyration ($r_g$) | | normal | 0.046405 | normal |
| | | lognormal | 0.065613 | |

**Supplementary Table S2 | The KS test between actual and controlled results.** $p$ equals zero, meaning it is very small, smaller than 1e-323. The KS test results indicate that the difference between actual and controlled results is significant.

| indicator \ case | | K-S test on empirical with (statistics, p-value) | | | |
|---|---|---|---|---|---|
| | | replaced | rigorously replaced | randomized frequency | shuffled |
| Fig.2e | $\Delta r$ calculated with the centroid of cells | (0.50, 0.0) | (0.31, 0.0) | | |
| Fig.2f | $r_g$ calculated with the centroid of cells | (0.85, 0.0) | (0.59, 0.0) | | |
| Fig.3a | the distribution of visitation frequency | $(0.25, 1.3 \times 10^{-4})$ | | $(0.34, 2.5 \times 10^{-8})$ | |
| Fig.3e | $k_f$ | (0.61, 0.0) | | | |
| Fig.3f | $\Delta r$ in exploration | | | | (0.063, 0.0) |

**Supplementary Table S3 | The Wasserstein Distance and MAPE between the real and model results.** We measure the performance of three models. On one hand, we use the Wasserstein Distance to measure the distribution distance between simulated probability distributions and the real distributions. Therefore, we calculate the Wasserstein Distance between the actual and simulated distribution of $r_g$ and $\Delta r$. On the other hand, regarding the $K$-frequency distribution, $r_g$ growth and $S$ growth, which are sequence data, we measure the difference between the actual and simulated value with MAPE (Mean Absolute Percentage Error). The difference between the results generated by the BEE model and the real results is much smaller, compared with the two null models.

| | Wasserstein Distance | | MAPE | | |
|---|---|---|---|---|---|
| Experiment | (b) $\Delta r$ | (c) $r_g$ | (d) $K$ frequency | (e) $r_g$ growth | (f) $S$ growth |
| Real & BEE | **0.700** | **0.740** | **0.069** | **0.255** | **0.079** |
| Real & #1 null model | 2.588 | 1.862 | 0.218 | 0.986 | 0.077 |
| Real & #2 null model | 0.861 | 1.654 | 1.307 | 0.911 | 1.932 |

**Supplementary Table S4 | The algorithm parameter setting.**

| Parameter | Doc2vec | UMAP | |
|---|---|---|---|
| | vector dimension | number of neighbors | metric |
| original | 200 | 15 | cosine |
| Experiment 1 | 100 | 15 | cosine |
| Experiment 2 | 200 | 50 | cosine |
| Experiment 3 | 200 | 15 | euclidean |